\documentclass[%
reprint,
superscriptaddress,
nofootinbib,
 amsmath,amssymb,
 aps,
]{revtex4-2}

\usepackage{graphics,epsfig,psfrag}
\usepackage{color}
\usepackage{xcolor}
\usepackage{graphicx}
\usepackage{dcolumn}
\usepackage{bm}
\usepackage{hyperref}

\usepackage{mathrsfs}

\hypersetup{
           breaklinks=true,   
           colorlinks=false,   
           pdfusetitle=true,  
        }

\begin{document}

\preprint{APS/123-QED}

\title{Overlapping of photon rings in black hole imaging}

\author{Oleg Yu.~Tsupko}
\email{tsupkooleg@gmail.com}
\affiliation{ZARM, University of Bremen, 28359 Bremen, Germany}

\author{Fabio Aratore}
\email{faratore@unisa.it}
\affiliation
{Dipartimento di Fisica “E.R. Caianiello”, Università degli studi di Salerno, Via Giovanni Paolo II 132, I-84084 Fisciano SA, Italy}
\affiliation{Istituto Nazionale di Fisica Nucleare (INFN), Sezione di Napoli -- Gruppo collegato di Salerno, Via Cintia, 80126 Napoli NA, Italy}

\author{Volker Perlick}
\email{volker.perlick@uni-bremen.de}
\affiliation{University of Bremen, Faculty 1, 28359 Bremen, Germany}

\date{\today}

\begin{abstract}
In this paper, we investigate the overlapping of photon rings --- higher-order images of a black hole’s luminous environment, concentrated near the shadow boundary and expected to be resolved in future observations. We consider a broad class of static spherically symmetric spacetimes and geometrically thin equatorial accretion disk with a prescribed inner radius and infinite outer extent, viewed by a polar observer. Depending on the inner radius of the disk, the thickness of each photon ring varies, and the rings may or may not overlap. By overlapping, we mean that portions of images appear at the same angular position on the observer’s sky. To characterize the overlapping, we introduce the radius of merging --- the value of the disk’s inner radius at which two photon rings of given orders begin to overlap. Since each radius of merging is labeled by two indices corresponding to the image orders, it becomes possible to arrange these radii in the form of an infinite-dimensional matrix where only the  upper right-hand corner is filled. This matrix, which we call the ``matrix of merging'', is a signature of spacetime only, and, once known, it provides a qualitative understanding of the overlapping pattern for any chosen value of the inner radius of the disk. Remarkably, the matrix of merging exhibits several universal properties that hold for all spherically symmetric metrics and can be established even without explicit calculation of light trajectories. Based on these properties, we demonstrate that certain overlapping patterns are universally forbidden across all such spacetimes and for any inner radius of the disk. Examples for the Schwarzschild and Reissner--Nordstr{\"o}m black holes are provided. The main application of our study is constraining the spacetime metric and the accretion model using observed photon ring overlaps.
\end{abstract}


\maketitle

\tableofcontents

\newpage

\section{Introduction}
\label{sec:introduction}

Supermassive black holes at the centers of galaxies can now be imaged via their shadows --- dark silhouettes appearing against bright background emission \cite{Falcke-2000, Bronzwaer-Falcke-2021, Cunha-Herdeiro-2018, Perlick-Tsupko-2022}. Recently, such shadows have been captured for the first time at the centers of M87 and the Milky Way \cite{akiyama2019first1, akiyama2019first2, akiyama2019first3, akiyama2019first4, akiyama2019first5, akiyama2019first6, Kocherlakota-2021, EHT-SgrA-2022-01, EHT-SgrA-2022-02, EHT-SgrA-2022-03, EHT-SgrA-2022-04, EHT-SgrA-2022-05, EHT-SgrA-2022-06}. Following this major success, black hole imaging has rapidly become one of the most important directions for studying black holes and distinguishing them from other ultracompact objects.

Planned future observational projects are designed to resolve finer substructures in black hole images \cite{johnson2020universal, pesce2021toward, Johnson-2023-Galaxies, Johnson-2024-BHEX, Ayzenberg-2025-review, Zhang-2025-future-observations}. In particular, it is expected that higher-resolution observations may reveal thin, bright rings — known as photon rings — around the boundary of the black hole shadow. These rings are lensed images of the black hole’s luminous environment, formed by photons that orbit the black hole multiple times before reaching the observer. Since these photons originate in the surrounding accreting matter and traverse regions of strong gravitational bending, photon rings encode information about both the central gravitating object and its environment, and can therefore be used to constrain them. Motivated by this, a comprehensive numerical and analytical study of photon rings has recently been carried out \cite{Gralla-2019, johnson2020universal, Gralla-Lupsasca-2020a-lensing, Gralla-Lupsasca-2020b-null-geodesics, Gralla-Lupsasca-2020c-shape-crit-curve, Gralla-Lupsasca-Marrone-2020, wielgus2021photon, Hadar-2021-photon-rings, Gan-Wang-2021-photon-ring, Guerrero-2022a-photon-rings, Guerrero-2022b-photon-rings, Broderick-2022-spin, BK-Tsupko-2022, Tsupko-2022-shape, Paugnat-2022-photon-rings-shape, Ayzenberg-2022-photon-rings, Carballo-Rubio-2022-photon-rings, Eichhorn-2023-photon-rings, da-Silva-2023-photon-rings, Papoutsis-2023-photon-rings, Staelens-2023-photon-rings, Broderick-Salehi-2023-photon-rings, Kocherlakota-2024a, Kocherlakota-2024b, Carballo-Rubio-2024-photon-rings, Deich-Yunes-2024-photon-rings, Cardenas-Avendano-2024, Aratore2024, Kobialko-2025, Frost-2025}. It is important to note that photon rings represent a particular type of higher-order images, a broad class of phenomena that have been studied extensively over the years, e.g., \cite{darwin1959gravity, Atkinson-1965, Luminet1979, Ohanian1987, Virbhadra-2000, bozza2001g, bozza2002gravitational, eiroa2002reissner, Perlick-2004-review, BK-Tsupko-2008, Bozza2010, stefanov2010connection, Tsupko-BK-2013, Tsukamoto-2016, aratore2021decoding, Aratore-Bozza-2024, Feleppa-Aratore-Bozza-2025}.

In this paper, we investigate the subject of photon rings overlapping. We consider a geometrically thin accretion disk in the equatorial plane of a spherically symmetric compact object, with an observer located on the symmetry axis at a large distance. We use an idealized model in which the disk is characterized by its inner and outer radii, with the outer radius taken to be infinite. We introduce the notion of a \textit{radius of merging} for two given images, defined as such inner radius of the accretion disk at which the two images start to merge. Namely, this occurs when the inner boundary of the lower-order ring (which is farther from the center) begins to merge with the outer boundary of the higher-order ring (which is closer to the center). Since the radius of merging is defined by two indices (image orders), it becomes possible to arrange the values of this radius in the form of an infinite-dimensional matrix where only the upper right-hand corner is filled. We refer to this matrix as to the \textit{matrix of merging}. The matrix of merging is a signature of the spacetime only, and, once known, it allows one to conclude whether or not overlap occurs for any chosen pair of rings and for any given value of the inner radius of the accretion disk.

Remarkably, the matrix of merging reveals several universal properties common to all spherically symmetric metrics, which can be established even without explicit calculation of light trajectories. We present a detailed analysis of the matrix properties and provide numerical examples for the Schwarzschild and Reissner--Nordström black holes. In particular, we demonstrate that some overlapping patterns (i.e., observational configurations in which some rings overlap while others remain separated) are universally forbidden across all spherically symmetric metrics and for any choice of the inner radius of the luminous disk. The potential application of our study consists in constraining the spacetime metric or accretion model based on the observed overlapping pattern of the photon rings.

The paper is organized as follows. In the next section, we introduce the notion of the radius of merging. Section~\ref{sec:matrix-merging} defines the matrix of merging, composed of these radii, and outlines its key properties. In Section~\ref{sec:light-deflection}, we discuss the calculation of light deflection and its relation to image order.  In Section \ref{sec:numerical}, the numerical procedure for determining the radii of merging and the matrix of merging is described. Section \ref{sec:example1-schw} and Section \ref{sec:example2-RN} present explicit examples of the matrix of merging for the Schwarzschild and Reissner--Nordstr{\"o}m metrics, respectively. In Section \ref{sec:constraining}, we explore how the observed overlapping pattern can be used to constrain either the spacetime geometry or the accretion disk model. Section \ref{sec:conclusions} provides a summary and conclusions.\\

\section{Radius of merging}
\label{sec:radius-merging}

In this section, we introduce the spacetime metric and the model under consideration, and introduce the key notion of this paper --- the radius of merging.

\subsection{Assumptions on the spherically symmetric metric}
\label{subsec:assumptions}

Let us begin by formulating the assumptions that we impose on the spacetime metric.

First, we consider a static and spherically symmetric spacetime for which the line element can be written in the form
\begin{equation}
ds^2 = - A(r) \, c^2 dt^2 + B(r) \, dr^2 + D(r)  \left( d \vartheta ^2 + \mathrm{sin} ^2 \vartheta \, d \phi ^2\right) \, .
 \label{eq:metric}
\end{equation}
We also demand the metric to be asymptotically flat or, in other words, that the metric coefficients $A(r)$, $B(r)$ and $D(r)$, for $r\to +\infty$, satisfy the following conditions: 
\begin{equation}
    A(r) \to 1 \, , \quad B(r) \to 1 \, , \quad \dfrac{D(r)}{r^2} \to 1 \, .
\label{eq:asy}
\end{equation}

Second, we assume that the potential
\begin{equation}
    V(r) = - \dfrac{D(r)}{A(r)}
\label{eq:V}
\end{equation}
has at least one extremum, meaning that the equation
\begin{equation}\label{eq:photon-sphere-equation}
    V'(r) = 0
\end{equation}
has at least one solution. Any such solution determines the radial coordinate of a photon sphere — a surface filled with circular orbits of light rays. A photon sphere is unstable with respect to radial perturbations if it is located at a local maximum of the potential \eqref{eq:V}, and stable if it is located at a local minimum. Unstable photon spheres have the property that light rays may asymptotically spiral toward them, whereas stable photon spheres allow light rays to oscillate about them.

We denote the outermost solution of Eq.~\eqref{eq:photon-sphere-equation} by $r_{\mathrm{ph}}$. Since our condition of asymptotic flatness \eqref{eq:asy} implies that the potential $V(r)$ goes to $- \infty$ for $r \to \infty$, the photon sphere at $r_{\mathrm{ph}}$ is necessarily unstable, at least with respect to perturbations in the positive $r$ direction. For our purposes, it is necessary to assume that the extremum of $V(r)$ at $r_{\mathrm{ph}}$ is not only a maximum but even an absolute maximum. This means that there may be additional unstable photon spheres, with stable ones in between, but at their locations the potential must remain smaller than $V(r_{\mathrm{ph}})$. This ensures that a light ray coming in from infinity cannot spiral towards any of these additional photon spheres. Note also that, from the above considerations, $V'(r)$ is negative for $r_{\mathrm{ph}} < r < \infty$, a fact we will use in Sec.~\ref{sec:light-deflection}.

The existence of an unstable photon sphere implies that light rays can complete an arbitrary number of revolutions around the center. Consequently, this results in the creation of infinitely many images of the same light source indexed by a number $n=0,1,2, ...$, called the order of the image. This index denotes how many times the photons intersect the optical axis, which is the straight coordinate line through the observer position and the center of the coordinate system. This definition is suitable only for spherically symmetric and static metrics. In this case it is consistent with the definition of the order, frequently used in recent papers, as ``the number of half-orbits'', see e.g. Ref.~\cite{johnson2020universal, Broderick-2022-spin, pesce2021toward, wielgus2021photon}.

Third, for our purposes it is crucial that there exists only one image of each order. Equivalently, light rays with different impact parameters must not produce the same deflection --- a condition satisfied in all physically relevant metrics. This assumption is equivalent to the monotonicity of the light deflection as a function of the impact parameter, which will be examined in more detail in Sec.~\ref{sec:light-deflection}.

For the purpose of this paper the existence of the photon sphere at $r_{\mathrm{ph}}$ is the most crucial assumption. Objects that possess a photon sphere are usually referred to as \emph{ultracompact objects} \cite{Iyer-1985, Cunha-Herdeiro-2018, Cardoso-2019}. Black holes, in particular, belong to this class. Correspondingly, the analysis in this paper is applicable not only to black holes but also to other ultracompact objects described by a spherically symmetric and static metric (\ref{eq:metric}) that satisfies the above-mentioned assumptions.

\subsection{Accretion disk and photon rings}
\label{subsec:photon-rings}

\begin{figure*}
    \centering
    \includegraphics[width=0.72\textwidth]{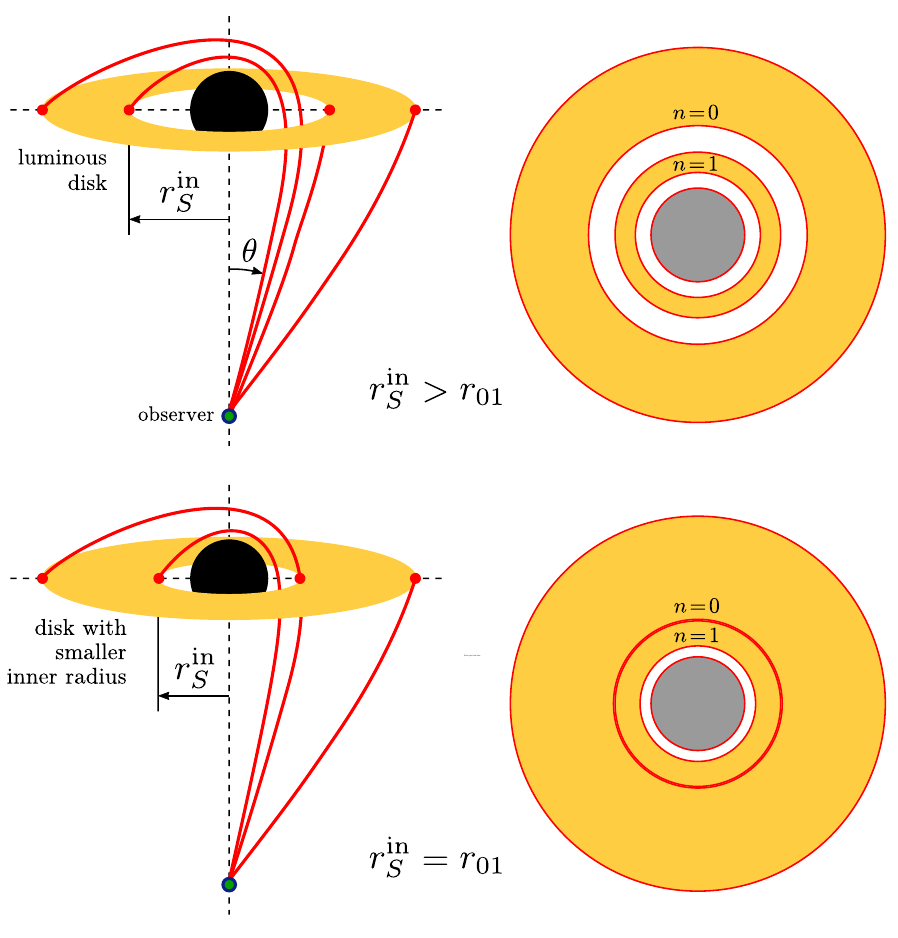}
    \caption{
Definition of the radius of merging (Subsec.~\ref{subsec:radius-merging-def}), illustrated through the example of merging primary and secondary images. The left panels show an equatorial accretion disk and a polar observer, along with the light rays that form the borders of primary and secondary images ($n = 0$ and $n = 1$ photon rings, respectively). The corresponding images, as they appear on the observer’s sky, are shown in the right panels.
In our simplified disk model, the angular thickness of each image, and therefore the overlap of images, depends only on the value of the inner radius of the accretion disk, $r_S^{\mathrm{in}}$.
In the upper panels, the inner radius $r_S^{\mathrm{in}}$ is large enough, such as the primary and secondary images are relatively thin and remain separated. If $r_S^{\mathrm{in}}$ is smaller, all images become thicker, since their inner boundaries shift closer to the center, while their outer boundaries remain fixed. In particular, the shrinking inner boundary of the primary image approaches the fixed outer boundary of the secondary image. The radius of merging, denoted $r_{01}$, is defined as the value of $r_S^{\mathrm{in}}$ at which the inner edge of the primary ($n=0$) image just touches the outer edge of the secondary ($n=1$) image. This configuration is shown in the lower panels.
Note that, for visualization purposes, the photon rings are not drawn to scale. In particular, throughout the paper, the outer radius of the disk is assumed to be infinite. The lower left panel demonstrates that the conditions (\ref{eq:mergingcondition-theta}) and (\ref{eq:mergingcondition}) cause the final segment of the ray forming the outer boundary of the secondary image to coincide with the trajectory of the ray forming the inner boundary of the primary image (see the text for more details).
}
    \label{fig:merging}
\end{figure*}

Let us now imagine that there is a geometrically thin accretion disk in the equatorial plane ($\vartheta=\pi/2$) of metric (\ref{eq:metric}). We consider a simple model in which the disk is prescribed by an inner radius $r_S^\mathrm{in}$ and an outer radius $r_S^\mathrm{out}$, see Fig.~\ref{fig:merging}. Moreover, for simplicity, the outer radius is assumed to be infinite: $r_S^\mathrm{out} \to \infty$. In such an approach, the only parameter defining the accretion model is the inner radius of the disk $r_S^\mathrm{in}$. An observer is located along the symmetry axis at a very large radial coordinate, i.e. in an asymptotically flat region.

The rays emitted by the luminous disk are lensed by the central object. Due to the existence of the photon sphere (see previous subsection), photons can circle around the object multiple times before reaching the observer. As a result, the observer sees an infinite sequence of lensed images of the disk known as \textit{photon rings} and enumerated by the order $n=0,1,2, ...$ as defined in Subsection \ref{subsec:assumptions}. We use the following convention:
\begin{itemize}
    \item  the primary image, or the $n=0$ photon ring;

    \item  the secondary image, or the $n=1$ photon ring;

    \item  the tertiary image, or the $n=2$ photon ring.
\end{itemize}
Throughout the paper, we use ``image'' and ``photon ring'' interchangeably to refer to a lensed image of the accretion disk.

Since the accretion disk represents an extended emission region bounded by inner and outer radii, and as we consider a polar observer, each image appears as a circular annulus with corresponding inner and outer angular radii on the observer's sky. All images are concentric and with increasing order $n$ they become thinner and thinner. In the limit $n \to \infty$ they converge upon the boundary of the shadow. The latter is defined as the locus on the observer’s sky corresponding to light rays that asymptotically approach the photon sphere \cite{Perlick-Tsupko-2022, Cunha-Herdeiro-2018}.

For the following reasoning it will be important to assume the accretion disk to be ``optically thin'', i.e. transparent. This ensures that images of all orders remain visible, even in cases when they overlap.

\subsection{Overlap of photon rings}
\label{subsec:overlap}

In this paper, we investigate the overlap of photon rings. We say that an image of order $n$ overlaps with an image of higher order $n'$ if portions of both images appear at the same angular position $\theta$ on the observer’s sky. Each successive photon ring is narrower than the previous one and is located closer to the shadow boundary. Therefore, overlap occurs when the inner radius of a lower-order image is smaller than the outer radius of a higher-order image. This
includes the case that a higher-order image is completely contained within a lower-order image. In our analysis, however, we will not distinguish the latter case from overlap in general.

In our simplified model, the overlap or lack thereof in a given metric is completely determined by the position of the inner boundary $r_S^\mathrm{in}$ of the disk. (Recall that the outer radius of the disk is fixed to be infinite.) Depending on the value of $r_S^\mathrm{in}$, the thickness of each photon ring varies, and therefore the photon rings may or may not overlap.

Because of our assumptions on the metric, for sufficiently large $r_S^\mathrm{in}$ all photon rings are thin enough to be separated from each other, showing a dark region between neighboring images. For example, in the Schwarzschild case all photon rings are separated from each other when $r_S^\mathrm{in}$ is equal to the radius of the innermost stable circular orbit ($6m$), see Fig.~7 in Ref.~\cite{BK-Tsupko-2022} and Fig.~5 in Ref.~\cite{Kocherlakota-2024b}.
If we begin at $r_S^{\mathrm{in}}=6m$, with decreasing $r_S^{\mathrm{in}}$ the inner boundary of each ring moves inward until it meets the outer boundary of the next ring, causing them to start merging. Depending on the value of $r_S^{\mathrm{in}}$, different combinations of overlapping and non-overlapping images can take place.

\subsection{Radius of merging}
\label{subsec:radius-merging-def}

In order to characterize the overlap of images (see previous subsection for details), we introduce the notion of the \textit{radius of merging}. The radius of merging $r_{nn'}$ is defined as the value of the inner radius of the accretion disk $r_S^{\mathrm{in}}$ at which a pair of chosen images of orders $n$ and $n'$ begin to merge. Here, ``begin to merge'' means that if the inner radius were to be decreased further, the two images would overlap. According to this definition, when $r_S^{\mathrm{in}}= r_{nn'}$, the inner boundary of the image of order $n$ and the outer boundary of the image of the higher order $n' > n$ have the same angular size on the observer's sky. Therefore, in order to find the radius of merging, we take $\theta_n(r_S^{\mathrm{in}})$, substitute $r_S^{\mathrm{in}}= r_{nn'}$ and set it equal to $\theta_{n'}(r_S^{\mathrm{out}})$, where $r_S^\mathrm{out} \to \infty$. Thus, the radius of merging $r_{nn'}$ can be determined from the following condition:
\begin{equation}
\theta_n (r_{nn'})= \underset{r_S^\mathrm{out} \to \infty}{\mathrm{lim}}\theta_{n'}(r_S^\mathrm{out}) \, , \quad  n \ge 0 \, , \; n' > n \, .
\label{eq:mergingcondition-theta}
\end{equation}
The angle $\theta$ is measured with respect to the radial direction (Fig.~\ref{fig:merging}).

The condition of asymptotical flatness \eqref{eq:asy} allows us to switch from the angular variable $\theta$ to the impact parameter $b$, cf. Eq.~(20) of Ref.~\cite{Aratore2024}. Namely, if a light ray with the impact parameter $b$ reaches an observer at a very large radial coordinate $r_O$, the incoming light ray will make an angle 
\begin{equation} \label{eq:theta-b-rO}
    \theta = \frac{b}{r_O}
\end{equation}
with respect to the radial direction. So in this situation we can use the impact parameter as a measure for the position on the sky. Then the definition \eqref{eq:mergingcondition-theta} of the radius of merging becomes
\begin{equation}
b_n(r_{nn'}) = 
\underset{r_S^\mathrm{out} \to \infty}{\mathrm{lim}} b_{n'}  ( r_S^\mathrm{out} ) \, ,
\quad  n \ge 0 \, , \; n' > n \, .
\label{eq:mergingcondition}
\end{equation}
Whereas it is not usually possible to calculate the radius of merging $r_{nn'}$ analytically, it is always possible to calculate it numerically, for any given metric and any pair of image orders $n$ and $n'$. If the value $r_{nn'}$ is found, one can immediately conclude what will happen to these two images on the observer's sky for any chosen value of the disk's inner radius $r_S^\mathrm{in}$. Namely:
\begin{itemize}
    \item  if $r_S^\mathrm{in} > r_{nn'}$, then images of orders $n$ and $n'$ don't overlap (upper panels of Fig.~\ref{fig:merging});

    \item  if $r_S^\mathrm{in} = r_{nn'}$, then the inner boundary of the $n$-ring touches the outer boundary of the $n'$-ring (bottom panels of Fig.~\ref{fig:merging});

    \item  if $r_S^\mathrm{in} < r_{nn'}$, then images of orders $n$ and $n'$ overlap.\\
\end{itemize}
Note that knowledge of $r_{nn'}$ does not allow us to distinguish the case that the image of order $n'$ is completely contained in the image of order $n$ from overlap in general. For making such a distinction it would be necessary to also compare the inner boundaries of both images rather than only the inner boundary of one and the outer boundary of the other.

Conditions \eqref{eq:mergingcondition-theta} and \eqref{eq:mergingcondition} lead to one more important property of the radius of merging, illustrated in the lower panels of Fig.~\ref{fig:merging} using the example of the $n=0$ and $n=1$ images. Let us consider the situation that the inner radius $r_S^\mathrm{in}$ equals the radius of merging $r_{01}$. Since the images touch at their boundaries, the light rays forming the outer boundary of the secondary image and the light rays forming the inner boundary of the primary image must subtend the same angular size on the observer’s sky [see Eq.~\eqref{eq:mergingcondition-theta}] and, according to Eq.~\eqref{eq:theta-b-rO}, must therefore share the same impact parameter [see Eq.~\eqref{eq:mergingcondition}]. Consequently, any particular ray from the outer boundary of the secondary image will coincide, along its final segment, with the trajectory of the corresponding ray from the inner boundary of the primary image, as illustrated in the lower left panel of Fig.~\ref{fig:merging}. In particular, this implies that the ray from the outer boundary of the secondary image intersects the equatorial plane precisely at the inner edge of the accretion disk, and this intersection point can be used to determine the radius of merging. In other words, if the images are initially separated and the inner radius of the accretion disk is gradually decreased, the merging of images occurs exactly when the disk's inner edge reaches the point where the ray from the outer boundary of the secondary image intersects the equatorial plane. This property will be further used in the next section.

It is important to emphasize the crucial difference between the dark annulus on the observer's sky that separates two images of an extended accretion disk from the ``gap'' as it was defined in our previous paper, Aratore \textit{et al.} \cite{Aratore2024}. In that earlier study we considered a circular ring at fixed $r_S$ as the source of emission. In that scenario, the images appeared as a series of concentric circular rings without a radial extension,  allowing us to define the ``gap'' as  the angular separation between two rings. This model could also be applied to an extended luminous disk as the light source, provided that the points of the images corresponding to the same emission radius $r_S$ were considered. One could choose, e.g., the points corresponding to the maximum of the emission (cf. Refs.~\cite{Broderick-2022-spin, wielgus2021photon}). In the present paper, we consider the emission coming from an accretion disk with given inner and outer boundaries, without specifying the emission profile. In this scenario, there is a dark region between the inner boundary of the image of order $n$ and the outer boundary of the image of order $n'$, as long as these two images do not overlap. We deliberately do not use the term ``gap'' for this dark region, to avoid confusion with our earlier paper. Obviously, in the case of an extended disk the inner boundary of one image and the outer boundary of another image are formed by rays emitted from different $r_S$.

\section{Matrix of merging: definition and properties}
\label{sec:matrix-merging}

In the previous section, we introduced the notion of the radius of merging, $r_{nn'}$, which characterizes the overlap of images of orders $n$ and $n'$. Recall that we are working within a broad class of spherically symmetric metrics of the form \eqref{eq:metric}, with all assumptions on the metric formulated in Subsection \ref{subsec:assumptions}.

Remarkably, as we will demonstrate below, the mutual relationship between the radii of merging is not arbitrary. Instead, there exist a number of universal properties that hold for the entire class of spacetimes under consideration.

In order to formulate these properties more systematically, it is convenient to organize the radii of merging into a matrix-like collection. Since each radius of merging is labeled by two indices, $n$ and $n'$ with $n' > n$, the entire set $\{ r_{nn'} \}$ can be naturally represented as a matrix where only the upper right-hand corner is filled,
\begin{equation}
    \begin{pmatrix}
        & \quad  r_{01} & r_{02} & r_{03}  & r_{04}  & r_{05}  & \dots\\
          & & r_{12}  & r_{13}  & r_{14}  & r_{15}  & \dots\\
          & &  & r_{23}  & r_{24}  & r_{25}  & \dots\\
          & &  &  & r_{34}  & r_{35}  & \dots \\
          & &  &  &  &  r_{45} &  \dots \\ 
          & & &  &  &  &  \vdots 
    \end{pmatrix} \ ,
\label{eq:rmatrix-general}
\end{equation}
which we refer to as the \textit{matrix of merging}.

To find the elements of this matrix (i.e., the radii of merging) for a specific spacetime, it is necessary to calculate the trajectories of light rays in that background. This procedure is discussed in detail in Sections~\ref{sec:light-deflection} and~\ref{sec:numerical}, and explicit examples of the matrix of merging are presented in subsequent Sections~\ref{sec:example1-schw} and~\ref{sec:example2-RN}.
However, a number of universal properties of the matrix of merging can be identified without explicit calculation of its elements. Below, we list and discuss these properties where we also include, at the end of the list, two properties which in contrast to the other ones are not clear without calculation. We include them here for completeness and will discuss them in later sections.

\begin{figure*}
    \centering
    \includegraphics[width=0.99\textwidth]{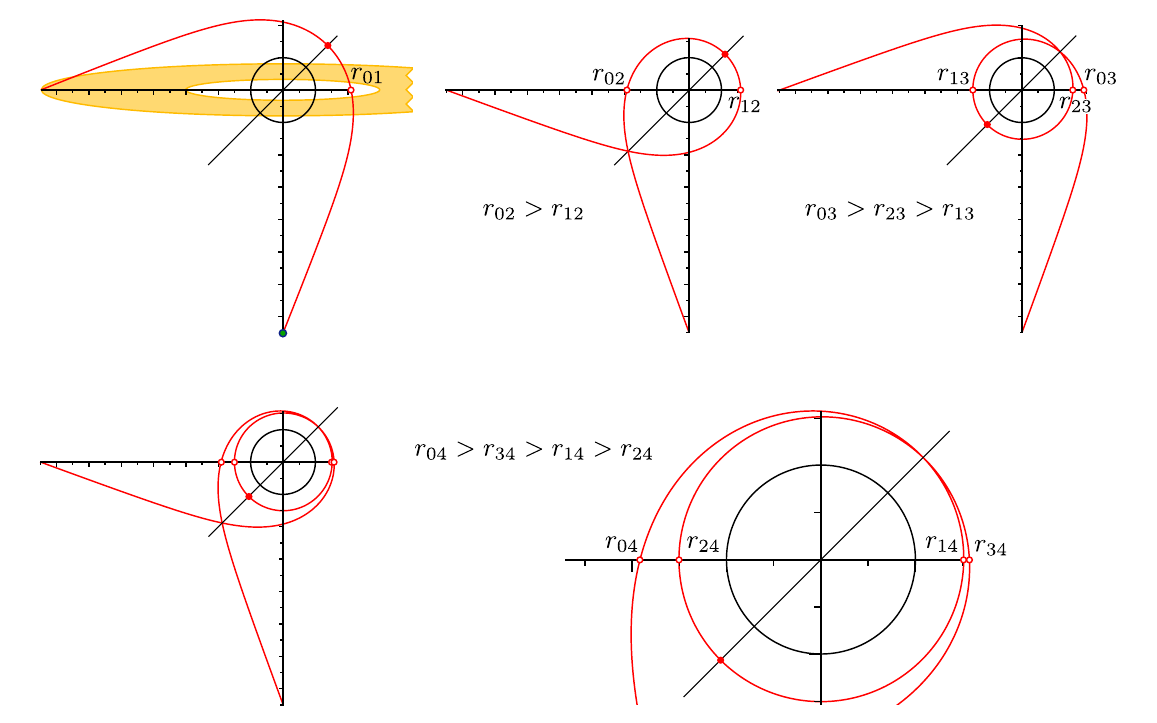}
    \caption{Graphical determination of the relationship between the elements of one column in the matrix of merging \eqref{eq:rmatrix-general}. Each panel shows a light ray emitted from the outer edge of the accretion disk and forming the outer boundary of an image of order~$n'$: $n'=1$ (upper left), $n'=2$ (upper middle), $n'=3$ (upper right), and $n'=4$ (lower panels, where the right panel is a zoom-in of the left). The radial coordinates of the points where the ray intersects the equatorial plane (marked with open circles) correspond to the radii of merging in the $n'$-th column of the matrix (see text for a detailed explanation). All trajectories are found numerically in the Schwarzschild metric. For visualization purposes, the outer radius of the accretion disk is set to $15m$ (instead of being infinite, as assumed in our model), and the observer is also placed at $15m$, ensuring symmetry of the ray paths. Due to the symmetry of each ray with respect to the diagonal line, the relationships between the matrix entries in a given column can be easily found based on their angular  ``separation'' from the point of closest approach, shown by the filled red dot. In particular, regardless of the column number $n'$, the value of $r_{0n'}$ is the largest in that column.}
    \label{fig:rays}
\end{figure*}

\begin{itemize}

\item 
After fixing the outer boundary of the disk (in our case at infinity) and the observer’s position (in our case at a large radial coordinate along the symmetry axis), the matrix depends solely on the metric. Consequently, it can be considered as a characteristic signature of the spacetime. While the answer to the question of whether the images of order $n$ and $n'$ overlap depends on both the spacetime metric and the inner radius of the luminous disk, the matrix of merging reflects the properties of the metric alone.

\item
As stated earlier, the determination of matrix elements requires the calculation of light ray trajectories. However, once the matrix is established, it allows us to immediately judge \textit{qualitatively} the overlapping and separation of images for any given value of the inner radius $r_S^{\mathrm{in}}$, without recalculation of geodesics. To assess overlap, it is sufficient to compare $r_S^{\mathrm{in}}$ with the matrix elements: if $r_S^{\mathrm{in}} > r_{nn'}$, the images of orders $n$ and $n'$ remain separate; if $r_S^{\mathrm{in}} < r_{nn'}$, the images overlap.

\item 
Knowledge of the matrix elements for a particular metric also allows us to rule out some overlapping patterns within that metric, based on the relative magnitudes of the radii of merging. Namely, if the element $r_{n''n'''}$ is smaller than $r_{nn'}$, then it is not possible to observe an overlapping pattern in which the images of orders $n''$ and $n'''$ overlap while the images of orders $n$ and $n'$ remain separated. Such a pattern cannot exist for any choice of the inner radius $r_S^{\mathrm{in}}$. The most interesting point here is that the relative magnitudes of the matrix elements across rows and columns exhibit a universal behavior, valid for all spherically symmetric metrics, as will be demonstrated below. Consequently, certain overlapping patterns can be ruled out not only for a particular metric, but universally for the entire class of spherically symmetric spacetimes considered here and for any choice of the inner radius of the accretion disk.

\item
Within each row of the matrix, the values decrease when the second index (which labels the columns) increases. For example,
\begin{equation} \label{eq:ineq-1}
r_{01} > r_{02} > r_{03} > ...    
\end{equation}
Indeed, when the inner boundary $r_S^\mathrm{in}$ of the disk decreases, the image of order $n$ first overlaps with the image of order $n+1$ and only afterwards with the image of order $n+2$.
This behavior follows from the monotonicity of the light deflection as a function of the impact parameter, which we required from the beginning. The outer boundary of an image approaches the boundary of the shadow monotonically when the order $n$ increases. This becomes more evident when looking at Fig.~\ref{fig:Schwarzschild} in Sec.~\ref{sec:example1-schw}: when calculating the outer boundaries of images, we deal only with the red branch of the shown curve, which decreases monotonically.

\item
All numbers in a given column of the matrix are determined by light rays with the same impact parameter. This non-obvious property becomes evident if one examines Eq.~\eqref{eq:mergingcondition}. On the right-hand side of this equation appears the impact parameter 
$b_{n'}$, which corresponds to the maximum possible angular radius of the $n'$-th image on the observer's sky and is determined only by the image order $n'$. This value remains unchanged  independently of the order $n$ of the image on the left-hand side of the equality \eqref{eq:mergingcondition}. Varying $n$ on the left while keeping $n'$ fixed on the right, we obtain all the elements in the $n'$-th column of the matrix. Hence, all these numbers correspond to rays with the same impact parameter.

This provides an alternative method for determining the entries in a given column of the matrix, as illustrated in Fig.~\ref{fig:rays}. Let us consider an image of order $n'$ and draw a ray that forms this image: namely, a ray emitted from the outer edge of the accretion disk. This ray fixes the outer boundary of the $n'$-th image on the observer's sky and the corresponding impact parameter. By definition \eqref{eq:mergingcondition}, to find the radius of merging $r_{nn'}$, we seek a ray with the same impact parameter that is emitted from the inner boundary of the disk and forms an image of lower order $n<n'$. Importantly, a light ray with the same impact parameter, but emitted from the equatorial plane, will follow the same geodesic and thus constitute a segment of the already traced trajectory. Therefore, each intersection of the original ray with the equatorial plane can be interpreted as a potential emission point at the inner boundary of the disk that generates a ray with the same impact parameter as the original ray (which produces the $n'$-th order image) but forms an image of lower order $n<n'$. Consequently, the intersection points where the ray crosses the equatorial plane correspond to the radii of merging in the $n'$-th column of the matrix: $r_{0n'}$, $r_{1n'}$, ... , $r_{nn'}$.

\item
Although the values in a given column do not increase or decrease monotonically, they exhibit a fixed relative ordering of magnitudes. Remarkably, this pattern is universal for all spherically symmetric metrics and can be obtained directly from the qualitative analysis of Fig.~\ref{fig:rays}, without explicit calculation of light-ray trajectories in specific metrics. First, we note that the ray's path is symmetric with respect to the line connecting the center of the black hole and the point of closest approach. Second, the radial coordinate increases monotonically away from the point of closest approach in both directions along the trajectory. Together, these properties make it straightforward to determine from the figure which radius of merging is larger or smaller, based on the angular ``separation'' between the intersection points with the equatorial plane and the point of closest approach. For example, Fig.~\ref{fig:rays} demonstrates that the following inequalities always hold:
\begin{equation} \label{eq:ineq-2}
    r_{12} < r_{02} \, ,
\end{equation}
\begin{equation}
    r_{13} < r_{23} < r_{03} \, ,
\end{equation}
\begin{equation}
    r_{24} < r_{14} < r_{34} < r_{04} \, .
\end{equation}

\item 
Independently of the column number $n'$, the element in the zeroth row, i.e., $r_{0n'}$, is the largest in that column. This conclusion also follows from the construction in Fig.~\ref{fig:rays}: the intersection point that determines $r_{0n'}$ is the farthest from the center.

This leads to the following conclusion: it is not possible to observe an overlap between the image of order $n'$ and any lower-order image with $0<n < n'$, without also observing an overlap between the $n'$-order image and the $n = 0$ image.

\item 
For the first three images ($n=0,1,2$) we have:
\begin{equation}
    r_{01} > r_{02} > r_{12}  \, .
\end{equation}
This follows directly from Eqs.~\eqref{eq:ineq-1} and \eqref{eq:ineq-2}. It leads to the following conclusion: it is not possible to observe an overlapping pattern in which the $n=1$ and $n=2$ rings overlap without the $n=0$ ring overlapping with both of them.

A collection of possible overlapping patterns of the first three images for different values of the accretion disk’s inner radius is shown in Fig.~\ref{fig:forbidden}, along with the forbidden configurations.

Additionally, this behavior will become evident in our examples for the Schwarzschild and Reissner–Nordström spacetimes; see Figs.~\ref{fig:LineplotShw}, \ref{fig:LineplotRN05}, and \ref{fig:LineplotRN1} in Sec.~\ref{sec:constraining}.

\item 
All elements of the matrix are greater than the photon-sphere radius $r_\mathrm{ph}$. For example, in the Schwarzschild case, all values exceed $3m$. This can be understood as follows. In Eq.~\eqref{eq:mergingcondition}, the right-hand side corresponds to the impact parameter of a ray forming the outer boundary of an image of order $n' \geq 1$. Such a ray starts from infinity and returns to infinity after bending around the black hole. It necessarily remains at radial coordinates larger than $r_\mathrm{ph}$. Since any possible emission point on the disk lies along this trajectory, it must also necessarily be larger than $r_\mathrm{ph}$.

Therefore, in the context of the merging behavior, there is no need to consider inner radii $r_S^{\mathrm{in}} \leq r_{\mathrm{ph}}$ --- no new overlaps will appear. Throughout this work, we restrict our analysis to $r_S^{\mathrm{in}} > r_{\mathrm{ph}}$.

\item

If $n' > n \ge 2$, the light deflection can be calculated with high accuracy using an analytical approximation known as the strong deflection limit \cite{darwin1959gravity, Luminet1979, Ohanian1987, bozza2001g, bozza2002gravitational, Dolan-2006, Bozza2010, Tsupko-2014, Feleppa-Bozza-Tsupko-2024, Feleppa-Bozza-Tsupko-2025}. This approximation provides an expression for the deflection angle that diverges logarithmically if the turning point of the light ray approaches the photon sphere. The strong deflection limit was originally formulated for the case that the light source is at infinity. In the case of photon rings, where the emission point is located in the vicinity of the black hole, we have to use the generalized version of the strong deflection limit for arbitrary source position which was derived by Bozza and Scarpetta \cite{bozza2007strong}, see also \cite{Bozza2010, Aldi-Bozza-2017}. We have applied these formulas for analytically studying the properties of higher-order photon rings in spherically symmetric spacetimes \cite{BK-Tsupko-2022, Tsupko-2022-shape, Aratore2024}. In particular, the apparent shape of higher-order images of a circular equatorial emission ring on the observer's sky was found by Tsupko \cite{Tsupko-2022-shape} for the Schwarzschild metric, and later generalized to arbitrary spherically symmetric spacetimes in Aratore \textit{et al.} \cite{Aratore2024}.

We will show in Subsection~\ref{subsec:schw-SDL} below that in the strong deflection limit the matrix element $r_{n(n+k)}$ depends only on $k$. For each $k$, these entries are situated on a diagonal line in the matrix of merging. As the strong deflection limit gives a good approximation for $n \ge2$, this means that from the third row onward all entries on such a diagonal are approximately the same.

\item

One more property has been revealed from our numerical calculations of the matrices of merging, presented in the following sections [see Eqs. ~\eqref{eq:Srmatrix}, \eqref{eq:RN05rmatrix}, \eqref{eq:RN1rmatrix}]. For completeness, it is also presented here.

We have observed that the values of the elements along each diagonal decrease monotonically. For example, one finds $r_{01} > r_{12} > r_{23} > ...$. While this behaviour holds for all our numerical examples and may potentially be universal for all other spherically symmetric spacetimes that satisfy our assumptions, we have not yet found a proof of this property for the general case.\\

\end{itemize}

\begin{figure*}
    \centering
    \includegraphics[width=0.97\textwidth]{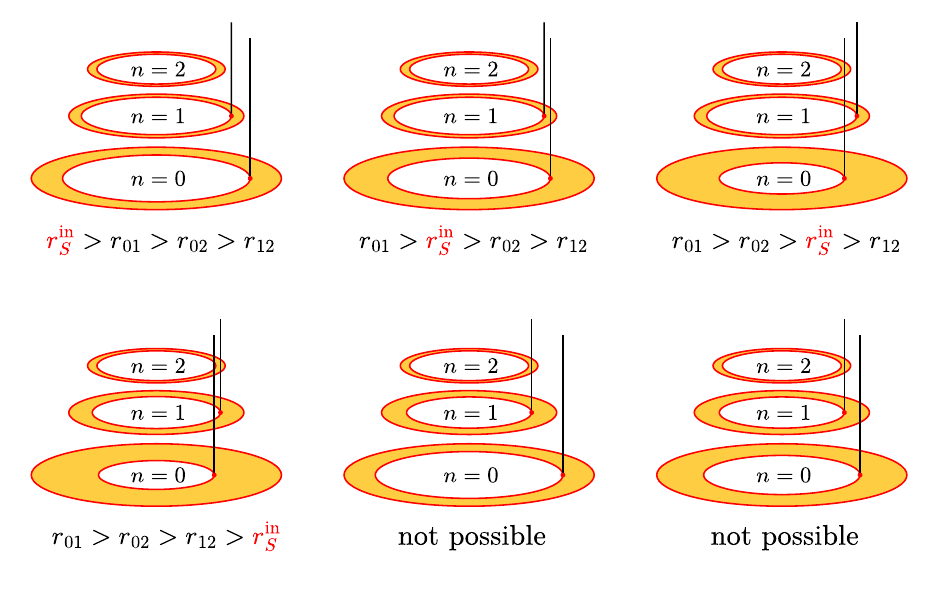}
    \caption{Overlapping patterns of the first three images for different inner radii of the accretion disk, including forbidden cases. Images are shown separately for clarity. Vertical lines indicate the inner boundaries of the $n=0$ and $n=1$ images, making it easier to identify overlaps visually. The upper row of panels shows overlapping patterns where $n=1$ and $n=2$ images are separated: (left) all images separated; (middle) $n=0$ image overlaps with $n=1$; (right) $n=0$ overlaps with both $n=1$ and $n=2$. The lower row of panels shows patterns where $n=1$ and $n=2$ overlap: (left) $n=0$ overlaps with both; (middle) $n=1$ and $n=2$ overlap, but $n=0$ does not overlap with $n=1$ and, consequently, not with $n=2$ --- forbidden; (right) $n=0$ overlaps with $n=1$ but not with $n=2$ --- forbidden.}
    \label{fig:forbidden}
\end{figure*}

\section{Calculation of light deflection, analysis of its monotonicity and relation to image order}
\label{sec:light-deflection}

In this section, we present the formulas for light deflection in terms of the change in the angular coordinate of a ray and analyze their monotonicity with respect to the impact parameter. As mentioned in Sec.~\ref{sec:radius-merging}, this monotonicity is crucial because it guarantees that of each order there is only one image. We also examine the relation between the light deflection and the image order $n$, and discuss the corresponding methods of computation.

\subsection{Calculation of light deflection}
\label{subsec:calculation-light-deflection}

Determining the orbit equation for lightlike geodesics in the spacetime \eqref{eq:metric} is a standard textbook exercise. One first observes that, because of the symmetry, every geodesic is contained in a coordinate plane through the origin. Consequently, the geodesics can be derived from the following Lagrangian: 
\begin{equation}
    \mathcal{L}(x, \dot{x} ) = \dfrac{1}{2} \left( - A(r) \, c^2 \dot{t}{}^2 + B(r) \, \dot{r}{}^2 + D(r) \, \dot{\tilde{\phi}}{}^2 \right) \, ,
\label{eq:Lagrangian}
\end{equation}
where $(r, \tilde{\phi} )$ are polar coordinates in the orbital plane and the overdot means derivative with respect to an affine parameter. It is important to distinguish $\tilde{\phi}$ in Eq.~\eqref{eq:Lagrangian} from $\phi$ in Eq.~\eqref{eq:metric}: the former represents the azimuthal angle in the orbital plane, while the latter is the azimuthal angle in the plane where we place the shining disk. This notation is also consistent with that used in our previous work \cite{Aratore2024}.

From Eq.~\eqref{eq:Lagrangian} we get two constants of motion,
\begin{equation}
    E = A(r) c^2 \dot{t} \, , \quad L = D(r) \dot{\tilde{\phi}} \, .
\end{equation}
Thereupon, the condition that the geodesic be lightlike gives us the orbit equation,
\begin{equation}
    \left( \dfrac{d r}{d \tilde{\phi}} \right) ^2 = \dfrac{D(r)^2}{b^2 A(r) B(r) } - \dfrac{D(r)}{B(r)} \, ,
\label{eq:orbit}
\end{equation}
where
\begin{equation}
    b^2 = \dfrac{c^2L^2}{E^2} \, .
\end{equation}
For light rays that come in from infinity, the constant $b$ equals the \textit{impact parameter}. Without loss of generality, we assume that $b \ge 0$. Equation \eqref{eq:orbit} can be rewritten in the form of an energy conservation law:
\begin{equation}
    \dfrac{b^2 B(r)}{D(r)} \left( \dfrac{dr}{d \tilde{\phi}} \right) ^2 + V(r) = - \, b^2 
\label{eq:con}
\end{equation}
with a potential $V(r)$ defined in Eq.~\eqref{eq:V}. Note that, according to \eqref{eq:con}, a light ray with impact parameter $b$ can exist only in the region where
\begin{equation}
    b^2 < - V(r) = \dfrac{D(r)}{A(r)} \, .
\label{eq:bV}
\end{equation}

By solving the orbit equation \eqref{eq:orbit} for $d\tilde{\phi}$ and integrating along the photon trajectory, we can find the angular shift $\Delta \tilde{\phi}$ experienced by a light ray emitted from a source at radial coordinate $r_S> r_{\mathrm{ph}}$ and detected by an observer at $r_O > r_S$. The resulting expression depends on whether the initial radial velocity $\dot{r}$ is positive or negative. Note that $\tilde{\phi}$ is a monotonic function of the affine parameter. Without loss of generality, we assume that $\Delta \tilde{\phi}$ is positive.

\begin{enumerate}
    \item[(i)]
If the initial radial velocity is negative, the angular shift is given by:
\begin{equation}
    \Delta \tilde{\phi} = 
     \label{eq:Deltaphiofb} 
\end{equation}
\[
\left( \int_{R}^{r_S} +\int_{R}^{r_O} \right) \sqrt{\frac{A(r)B(r)b^2}{D(r)^2-A(r)D(r)b^2}}\, dr \, .
\]
    In the case of Eq.~\eqref{eq:Deltaphiofb}, the photon initially moves inward, decreasing its radial coordinate until it reaches a turning point at a radius coordinate $R$. At this point of closest approach the radial velocity $\dot{r}$ vanishes and then changes sign; subsequently the photon moves outward toward the observer. This is possible only for values of $R$ that are bigger than the  radius of the photon sphere $r_{\mathrm{ph}}$. For $R \to r_{\mathrm{ph}}$ the angular shift $\Delta \tilde{\phi}$ diverges.

The impact parameter $b$ is related to the radius coordinate $R$ of the point of closest approach by the equation
\begin{equation}
    b^2=\frac{D(R)}{A(R)} = - V(R) 
    \label{eq:impactpar}
\end{equation}
which implies that 
\begin{equation}
    2b \dfrac{db}{dR}= - V'(R) \, .
    \label{eq:VpR}
\end{equation}
As we assume (recall Subsec.~\ref{subsec:assumptions}) that $V'(R)$ is strictly negative for all possible values $r_{\mathrm{ph}} < R < \infty$, this equation makes sure that $b$ is a monotonically increasing function of $R$. If $R$ goes from $r_{\mathrm{ph}}$ to $r_S$, $b$ goes from $b_{\mathrm{cr}}$ to $\sqrt{-V(r_S)}$, where the critical impact parameter $b_{\mathrm{cr}}$ is defined by
\begin{equation}
    b^2_{\mathrm{cr}} = \frac{D(r_{\mathrm{ph}})}{A(r_{\mathrm{ph}})} = - V(r_{\mathrm{ph}}) \, .
\end{equation}

\item[(ii)] If the initial radial velocity is positive, the angular shift is given by:
\begin{equation}
\Delta \tilde{\phi} = \int_{r_S}^{r_O}\sqrt{\frac{A(r)B(r)b^2}{D(r)^2-A(r)D(r)b^2}}\, dr \, .
\label{eq:Deltaphiofbnoinversion}
\end{equation}
In the case of Eq.~\eqref{eq:Deltaphiofbnoinversion}, the photon travels outward from the source to the observer, with a monotonically increasing radial coordinate.

While in Eq.~\eqref{eq:Deltaphiofb} the condition $b > b_{\mathrm{cr}}$ always holds, in Eq.~\eqref{eq:Deltaphiofbnoinversion} the impact parameter can be either larger or smaller than $b_{\mathrm{cr}}$. This divides the application of Eq.~\eqref{eq:Deltaphiofbnoinversion} into two subcases.

\begin{enumerate}

\item[(ii.a)]
If $b > b_{\mathrm{cr}}$, Eq.~\eqref{eq:Deltaphiofbnoinversion} describes the case where the path from $r_S$ to $r_O$ is part of a longer trajectory with a turning point: tracing the ray backward in time, it passes through this turning point before returning to larger $r$. In this case, the value of $R$ exists, can be computed using Eq.~\eqref{eq:impactpar}, and can be substituted in place of $b$ (see Sec.~\ref{sec:numerical}).

\item[(ii.b)]
If $b < b_{\mathrm{cr}}$, however, no point of closest approach exists, not even when the ray is maximally extended. In this case $R$ cannot be defined. (Here we make use of our assumption that the potential $V$ has an \emph{absolute} maximum at $r_{\mathrm{ph}}$. Therefore, a light ray traveling inward with impact parameter $b < b_{\mathrm{cr}}$ cannot return to infinity.) In this situation, the calculations must be performed directly in terms of $b$ using Eq.~\eqref{eq:Deltaphiofbnoinversion} (see discussion in Sec.~\ref{sec:constraining}).

\end{enumerate}
\end{enumerate}

\subsection{Monotonicity of light deflection}
\label{subsec:monotonicity}

For our purpose it is of crucial relevance to make sure that rays with different impact parameters cannot have the same angular shift $\Delta \tilde{\phi}$. To that end we have to investigate the monotonicity of the two branches of $\Delta \tilde{\phi}$ as a function of $b$, where one branch is given by \eqref{eq:Deltaphiofb} and the other one by \eqref{eq:Deltaphiofbnoinversion}.

In the case of \eqref{eq:Deltaphiofbnoinversion}, which applies to light rays which have no turning point between the source and the observer, this is easy. If we keep $r_S$ and $r_O$ fixed, with $r_{\mathrm{ph}} < r_S < r_O$, Eq.~\eqref{eq:Deltaphiofbnoinversion} implies 
\begin{equation}
  \dfrac{d}{db}\Delta \tilde{\phi} 
  = \int_{r_S}^{r_O}\sqrt{\dfrac{B(r)}{D(r)}} 
   \, \dfrac{\big(-V(r) \big) dr}{\big(-V(r)-b^2 \big)^{3/2}} \, . 
\end{equation}
As the right-hand side is manifestly positive [recall (\ref{eq:bV})], $\Delta \tilde{\phi}$ is monotonically increasing on the entire interval on which this branch is defined, i.e., for $0 < b < \sqrt{-V(r_S)}$. For the Schwarzschild spacetime this branch is plotted as a dashed (blue) curve in Fig. \ref{fig:Schwarzschild} of Sec.~\ref{sec:example1-schw}.

The situation is more complicated in the case of \eqref{eq:Deltaphiofb}. In this case we have to express $b$ in terms of $R$ with the help of (\ref{eq:impactpar}). 
Then Eq.~\eqref{eq:Deltaphiofb} can be rewritten as
\begin{equation}
\Delta \tilde{\phi} = \left( \int_{R}^{r_S} +\int_{R}^{r_O} \right) 
\sqrt{\frac{B(r)}{D(r)}} \dfrac{dr}{\sqrt{\frac{V(r)}{V(R)} - 1 }} \, .
\label{eq:DeltatphiR}
\end{equation}
According to (\ref{eq:VpR}), $\Delta \tilde{\phi}$ is a monotonic function of $b$ if and only if it is a monotonic function of $R$, i.e., if and only if the derivative of the right-hand side of (\ref{eq:DeltatphiR}) with respect to $R$ has no zeros. Demonstrating that this is true is subtle because a straight-forward calculation of the derivative gives an undetermined expression of the form $-\infty + \infty$,
\[
\dfrac{d}{dR} \Delta \tilde{\phi} = 
- \underset{r \to R}{\mathrm{lim}}  
\sqrt{\frac{B(r)}{D(r)}} \dfrac{2}{\sqrt{\frac{V(r)}{V(R)} - 1 }} 
\]
\begin{equation}
+ \, \dfrac{V'(R)}{2 V(R)^2}  
\left( \int_{R}^{r_S} +\int_{R}^{r_O} \right)
\sqrt{\dfrac{B(r)}{D(r)}} \dfrac{V(r) dr}{\left(\frac{V(r)}{V(R)} - 1 \right)^{3/2}}
\, .
\label{eq:dDphiR}
\end{equation}
Here $b$ takes values between $b_{\mathrm{cr}}$ and $\sqrt{-V(r_S)}$ and, correspondingly, $R$ takes values between $r_{\mathrm{ph}}$ and $r_S$. Whereas it is always true that $\Delta \tilde{\phi}$ goes to infinity for $b$ to $b_{\mathrm{cr}}$ (i.e., $R$ to $r_{\mathrm{ph}}$) and to a finite positive value for $b$ to $\sqrt{-V(r_S)}$ (i.e., $R$ to $r_S$) where the other branch is met, it is not in general true that $\Delta \tilde{\phi}$ is a \textit{monotonically} decreasing function of $b$ (or $R$) on this interval. We are not aware of any physically relevant metric in which this monotonicity property is violated. As we will prove in Sec.~\ref{sec:example2-RN} below, it holds in particular for the Reissner-Nordstr{\"o}m metric, which contains the Schwarzschild metric as a special case. However, we demonstrate in the next Subsection that it is possible to construct (contrived) examples which satisfy all our other assumptions but \emph{do} violate the monotonicity property. As the monotonicity property is crucial for our following reasoning, it is necessary to check for each individual metric whether it is satisfied.

\subsection{Example of monotonicity violation}
\label{subsec:nonmonotonic}

As indicated above, the condition of asymptotic flatness does not in general imply that there is exactly one image of each order $n$. To exemplify this statement, we give here a (contrived) example where three images of order 1 exist:
\begin{equation}
A(r) =  1- \frac{2m}{r} \, , \; 
B(r) =  1- \dfrac{8}{9} \, \mathrm{sin} \Big( 500 \dfrac{m}{r} \Big) \, , \; 
D(r) = r^2 \, .
\label{eq:Example}
\end{equation}
This spacetime is asymptotically flat, and the potential $V(r)$ is the same as in the Schwarzschild spacetime; in particular, there is still an (outermost) unstable photon sphere located at $r=3m$, and the potential $V(r)$ is monotonically decreasing from $V(3m) = - 27$ to $V(\infty) = - \infty$ . 

\begin{figure}[ht]
    \centering
    \includegraphics[width=0.48\textwidth]{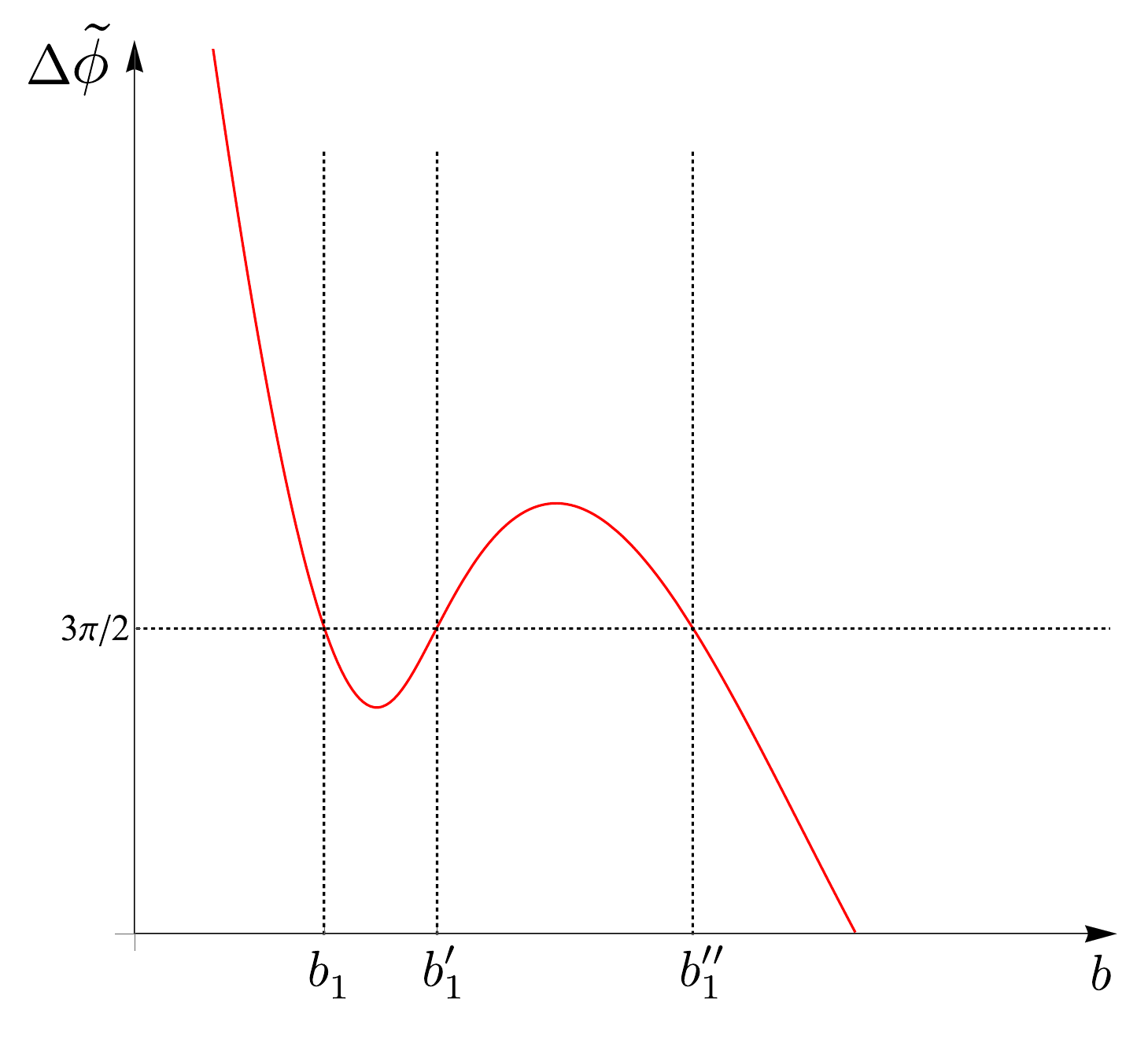}
    \caption{$\Delta \tilde{\phi}$ as a function of the impact parameter $b$ for the metric (\ref{eq:Example}). The source is at $r_S = 6m$, $\vartheta _S = \pi/2$ and the observer is at $r_O = \infty$, $\vartheta _O = \pi$. Then there are three images of order 1 whose impact parameters $b_1$, $b'_1$ and $b''_1$ are marked in the picture.}
    \label{fig:Example}
\end{figure}

However, the oscillatory behavior of the metric coefficient $B(r)$ produces a non-monotonic behavior of the angular shift $\Delta \tilde{\phi}$, see Fig.~\ref{fig:Example}. If we assume that the angular positions of source and observer are chosen as for the construction of our matrix, an image of order $n$ corresponds to an angular shift of $\Delta \tilde{\phi} = (n+1/2) \pi$. We read from the picture that then there are three different images of order $n=1$.

By making the oscillatory behaviour of $B(r)$ even stronger, one could produce spacetimes where images of arbitrarily high order are non-unique.

\subsection{Relation of light deflection to image order}
\label{subsec:image-order-and-deflection}

In this paper, we consider only the case of a polar observer. This configuration allows for an easy identification of the azimuthal shift corresponding to an image of order $n$ (see Fig.~\ref{fig:merging}):
\begin{equation}
     \Delta \tilde{\phi} = \left( n + \frac{1}{2} \right) \pi \, .
 \label{eq:n}
\end{equation}
$\Delta \tilde{\phi}$ has to be calculated as a function of $b$ with  \eqref{eq:Deltaphiofb} or \eqref{eq:Deltaphiofbnoinversion}. Light rays with $b \le b_{\mathrm{cr}}$ cannot have a turning point. In the notation of Subsection \ref{subsec:calculation-light-deflection} this is the case (ii.b), so we have to use \eqref{eq:Deltaphiofbnoinversion}. Light rays with $b > b_{\mathrm{cr}}$ do have a turning point at some minimum radius $R$, where $b$ and $R$ are related by \eqref{eq:impactpar}. We have to distinguish the case that the point of closest approach is inside the segment of the light ray between the light source and the observer from the case that it is outside. In the first case, which in the notation of Subsec. \ref{subsec:calculation-light-deflection} is the case (i), we have to use \eqref{eq:Deltaphiofb}, in the second case [case (ii.a)] we have to use \eqref{eq:Deltaphiofbnoinversion}.
For each fixed $r_S > r_{\mathrm{ph}}$ and $r_O = \infty$ this gives us $ \Delta \tilde{\phi}$ as a function of $b$. This function consists of two branches, as outlined in Subsec. \ref{subsec:monotonicity}. Whereas our general assumptions guarantee that the lower branch is monotonic, this is not necessarily true for the upper branch, as we have exemplified in Subsec. \ref{subsec:nonmonotonic}. In any case, $\Delta \tilde{\phi}$ goes to $\infty$ with $b \to b_{\mathrm{cr}}$ or $R \to r_{\mathrm{ph}}$. If the monotonicity property is satisfied, inserting $\Delta \tilde{\phi}$ as a function of $b$ into (\ref{eq:n}) gives a unique value of $b$ (or of $R$ in all cases where the latter is defined) for each order $n$. For the Schwarzschild metric, these values are marked by horizontal lines in Fig.~\ref{fig:Schwarzschild} of Sec.~\ref{sec:example1-schw} for $n=0,1,2,3,4$.

\subsection{Parts of images inside the shadow region}
\label{subsec:images-inside-shadow}

The existence of a critical value of the impact parameter also implies that a black hole illuminated by a bright background will cast a shadow. For an observer at a large radius coordinate $r_O$, the angular radius of the shadow equals $b_{\mathrm{cr}} / r_O$.

In the present paper, we consider a black hole surrounded by a luminous accretion disk lying in the equatorial plane, with the observer situated along the polar axis. It should be emphasized that in this configuration the shadow region is not necessarily completely dark: light rays issuing from the disk with $b < b_{\mathrm{cr}}$, which cannot have turning points, may reach the observer and thus create images inside the shadow. This happens if the radius coordinate of the inner boundary of the luminous disk is sufficiently small (though not necessarily smaller than the photon sphere radius); see the figures in Sec.~\ref{sec:constraining}. In this subsection, we discuss this effect in more detail.

As we assume that the disk extends to infinity and that the observer is also at infinity, the outer boundaries of all images always lie outside the shadow on the observer’s sky. Indeed, the outer boundaries of all images are formed by rays that originate from infinity, necessarily pass through a point of closest approach, and then return to infinity. All such rays satisfy $b > b_\mathrm{cr}$ and $R > r_\mathrm{ph}$, and therefore contribute to image points outside the shadow. This also means that in our calculations of all radii of merging (Sec.~\ref{sec:numerical}), we deal exclusively with rays for which $b > b_\mathrm{cr}$. However, when plotting images for a specific $r_S^{\mathrm{in}}$ (Sec.~\ref{sec:constraining}), some rays may have $b < b_\mathrm{cr}$, causing the inner parts of images to appear inside the shadow.

So we emphasize that, even though we only consider sources outside the photon sphere (i.e., $r_S^{\mathrm{in}} > r_{\mathrm{ph}}$, see the discussion in Sec.~\ref{sec:matrix-merging}), it is still possible for the inner boundary of an image to lie inside the shadow.

If a ray has $b < b_{\mathrm{cr}}$ and begins with $\dot{r} > 0$ [case (ii.b) in Subsec.~\ref{subsec:calculation-light-deflection}], it travels outward monotonically without encountering a turning point and contributes to an image inside the shadow. This is possible for images of any order, even at high order, provided the inner radius of the accretion disk, $r_S^{\mathrm{in}}$, is sufficiently small. In particular, if $r_S^{\mathrm{in}}$ is close to $r_{\mathrm{ph}}$, such a ray may spiral around the black hole multiple times while monotonically increasing its radial coordinate, before finally reaching the observer. The angular shift can be calculated using Eq.~\eqref{eq:Deltaphiofbnoinversion}. We can follow such a ray backwards in time until it reaches the horizon. Of course, it can cross the horizon only if we consider a white-hole extension of the spacetime.

If instead the ray has $b > b_{\mathrm{cr}}$ and $\dot{r} > 0$ [case (ii.a) in Subsec.~\ref{subsec:calculation-light-deflection}], then it reaches the observer after bending around the black hole and forms part of an image outside the shadow. Although the ray moves outward toward the observer, tracing it back in time reveals that it reaches a turning point before returning to larger $r$. The angular shift can be calculated using Eq.~\eqref{eq:Deltaphiofbnoinversion}. Rays with $b > b_{\mathrm{cr}}$ and initial $\dot{r} < 0$ (directed inward at emission, case (i) in Subsec.~\ref{subsec:calculation-light-deflection}) may also reach the observer after passing through a turning point, and again contribute to images outside the shadow. The angular shift can be calculated using Eq.~\eqref{eq:Deltaphiofb}.

For completeness, we briefly comment on the case where a source lies inside the photon sphere, which is not considered in our calculation of the matrix of merging. Any observable image of such a source must be formed by light rays with an impact parameter $b < b_{\mathrm{cr}}$. These rays always move outward ($\dot{r} > 0$) and do not encounter a turning point. The resulting images appear within the shadow. If such a ray is traced in reverse time (from the observer back toward the black hole), it would be captured without turning. Note that light rays emitted inside the photon sphere with $b = b_{\mathrm{cr}}$ asymptotically approach the photon sphere from the inside, while those with $b > b_{\mathrm{cr}}$ are inevitably captured (e.g., Ames and Thorne \cite{Ames-Thorne-1968}).

The observation that appropriately placed light sources may produce images inside the shadow was emphasized by Dokuchaev and Nazarova \cite{DokuchaevNazarova2019}; see also Chael \textit{et al.} \cite{Chael-2021}.\\

\section{Numerical calculation of radius of merging and matrix of merging}
\label{sec:numerical}

This section is specifically dedicated to the numerical procedure used to determine the radius of merging $r_{n n'}$ defined in Sec.~\ref{sec:radius-merging}. The collection of all $r_{nn'}$ forms the matrix of merging, as defined in
Sec.~\ref{sec:matrix-merging}.

As discussed above in Subsec.~\ref{subsec:images-inside-shadow}, when calculating radii of merging, we deal only with the rays with $b > b_{\mathrm{cr}}$. It means that only cases (i) and (ii.a) from the Subsec.~\ref{subsec:calculation-light-deflection} will be used in this section. By substituting the relation \eqref{eq:impactpar} between the impact parameter $b$ and the closest-approach distance $R$ into Eqs.~\eqref{eq:Deltaphiofb} and \eqref{eq:Deltaphiofbnoinversion}, we obtain equivalent expressions for the deflection $\Delta \tilde{\phi}$ in terms of $R$, more suitable for our analysis:
\begin{equation}
    \Delta \tilde{\phi} = \int_{R}^{r_S} g(r,R)\, dr + \int_{R}^{r_O} g(r,R) \, dr
    \label{eq:DeltaphiofR} 
\end{equation}
and
\begin{equation}
    \Delta \tilde{\phi} = \int_{r_S}^{r_O} g(r,R) \, dr \, .
    \label{eq:DeltaphiofRnoinversion}
\end{equation}
Here we have defined the function
\begin{equation}
    g(r,R) = \sqrt{\frac{A(r)B(r)D(R)}{A(R)D^2(r) - A(r)D(r)D(R)}} \, .
    \label{eq:grR}
\end{equation}
Equation \eqref{eq:DeltaphiofRnoinversion} may appear contradictory at first glance, because it applies when the photon's path does not go through an inversion point, yet the integrand is expressed in terms of the closest-approach distance $R$. However, as explained in Subsec. \ref{subsec:calculation-light-deflection}, when considering a light ray with $b > b_{\mathrm{cr}}$ that moves monotonically outward (i.e. with $\dot{r}>0$) from the source to the observer, one can always trace the trajectory backwards in time to identify a point of closest approach which lies at a radial coordinate smaller than that of the emission point $r_S$, see the case (ii.a) from Subsec.~\ref{subsec:calculation-light-deflection}. The case of a ray with $b< b_{\mathrm{cr}}$, which meets the horizon if traced backwards, will be discussed later in Sec.~\ref{sec:constraining}; see the case (ii.b) in Subsec.~\ref{subsec:calculation-light-deflection}.

As previously mentioned, for the outer border of the $n'$-th image to coincide with the inner border of the $n$-th image as seen by a distant observer, the condition for merging is that the corresponding light rays must reach the asymptotic region with the same impact parameter as given in Eq.~\eqref{eq:mergingcondition}. Using the relation between $b$ and $R$ from Eq.~\eqref{eq:impactpar}, this condition is equivalent to requiring that both light rays have the same value of the minimum radius $R$. The radius of merging $r_{n n'}$ is therefore found using a two-step numerical procedure.

Let us first consider radii of merging of the type $r_{0 n'}$, where the primary image begins to merge with higher-order ones of order $n'$. The inner boundary of primary image is formed by photons that increase monotonically in radial coordinate as they propagate from the source to the observer. In this case, Eq.~\eqref{eq:DeltaphiofRnoinversion} applies. For the outer boundary of a higher-order image, on the other hand, Eq.~\eqref{eq:DeltaphiofR} must be used.

The first step consists in solving for the value of $R$ using the condition that the light ray corresponding to the outer boundary of the $n'$-th image originates from spatial infinity ($r_S \to \infty$), reaches a distant observer at infinity ($r_O \to \infty$), and undergoes the angular shift given by Eq.~\eqref{eq:n}:
\begin{equation}
     \left(n'+\frac{1}{2} \right) \pi = 2 \int_{R}^{+\infty} g(r,R) \, dr \ .
     \label{eq:firststep}
\end{equation}
Once $R$ has been determined, it is substituted into Eq.~\eqref{eq:DeltaphiofRnoinversion}, which for the primary image takes the form:
\begin{equation}
    \frac{\pi}{2} = \int_{r_{0 n'}}^{+\infty} g(r,R) \, dr \ .
    \label{eq:secondstepprimary}
\end{equation}
Solving this equation yields the radius of merging $r_{0 n'}$.

In the case of merging between two images of higher order, with $n \neq 0$, the procedure remains conceptually the same. The first step is still Eq.~\eqref{eq:firststep}, which determines the shared value of $R$ for the two rays. Then, for the inner boundary of the image of order $n$, we first assume that the corresponding light ray passes through a turning point, and therefore use Eq.~\eqref{eq:DeltaphiofR} for calculation of the angular shift. Correspondingly, the radius of merging $r_{n n'}$ is obtained by solving:
\begin{equation}
        \left( n + \frac{1}{2} \right) \pi = \int_{R}^{r_{n n'}} g(r,R) \, dr + \int_{R}^{+\infty} g(r,R) \, dr \, .
    \label{eq:secondstep}
\end{equation}
In some cases, however, the value of $R$ found in the first step leads to a situation where Eq.~\eqref{eq:secondstep} has no solution for any $r_{n n'}$. This indicates that the light ray corresponding to the $n$-th image does not pass through a turning point and instead always propagates with $\dot{r} > 0$; see the case (ii.a) in Subsec.~\ref{subsec:calculation-light-deflection}. In such cases, the appropriate equation for finding the radius of merging becomes:
\begin{equation}
    \left( n + \frac{1}{2} \right) \pi = \int_{r_{n n'}}^{+\infty} g(r,R) \, dr \, .
     \label{eq:secondstepnoinversion}
\end{equation}
Such cases arise for specific combinations of $n$ and $n'$. For example, referring to Fig.~\ref{fig:rays}, when $n'=3$, the tertiary image ($n=2$) goes through the inversion point, and Eq.~\eqref{eq:secondstep} must be used. Meanwhile, the secondary image ($n = 1$) proceeds directly to the observer, always increasing in radial coordinate, and must therefore be treated with Eq.~\eqref{eq:secondstepnoinversion}.

It is also possible to formulate a general condition for determining whether a light ray associated with order $n$ reaches an inversion point, given a reference image of order $n'$. A light ray of order $n'$ experiences a deflection angle of $(n' + 1/2)\pi$. The point of closest approach is reached halfway along this trajectory, corresponding to a covered angle of $(1/2)(n' + 1/2)\pi$. Any light ray of order $n$ with a deflection angle smaller than this value will not reach a turning point. This leads to the inequality:
\begin{equation}
\left(n + \frac{1}{2}\right) \pi < \frac{1}{2} \left(n' + \frac{1}{2} \right) \pi \, ,
\end{equation}
which simplifies to
\begin{equation}
n < \frac{1}{2} \left( n' - \frac{1}{2} \right) \, .
\label{eq:inversioncondition}
\end{equation}
Consequently, when this condition is satisfied, Eq.~\eqref{eq:secondstepnoinversion} must be used. More explicitly:
\begin{itemize}
    \item If $n'$ is odd, then the light ray of order $(n'-1)/2$ is the first that does not reach the turning point (e.g., for $n'=3$, the $n=1$ ray monotonically increases in radius).
    \item If $n'$ is even, then the ray of order $n'/2$ is the last to reach the inversion point, and Eq.~\eqref{eq:secondstep} still applies.
\end{itemize}  
Moreover, the integer closest to the right-hand side of Eq.~\eqref{eq:inversioncondition} corresponds to the minimum value of the radius of merging $r_{nn'}$ in each column of the matrix of merging defined in Eq.~\eqref{eq:rmatrix-general}.

From the point of view of numerical integration, accuracy is improved by introducing the new variable
\begin{equation}
\eta = 1 - \frac{R}{r} \, ,
\label{eq:defeta}
\end{equation}
for some fixed $R > r_{\mathrm{ph}}$. Letting $\eta_S$ and $\eta_O$ be the values corresponding to the source and observer positions, respectively, Eqs.~\eqref{eq:DeltaphiofR} and \eqref{eq:DeltaphiofRnoinversion} are rewritten as
\begin{equation}
\Delta \tilde{\phi} = \int_{0}^{\eta_S} h(\eta,R) \, d\eta + \int_{0}^{\eta_O} h(\eta,R) \, d\eta \, ,
\label{eq:Deltaphiofeta}
\end{equation}
and
\begin{equation}
\Delta \tilde{\phi} = \int_{\eta_S}^{\eta_O} h(\eta,R) \, d\eta \, ,
\label{eq:Deltaphiofetanoinversion}
\end{equation}
where
\begin{equation}
h(\eta,R) = \frac{R}{(1 - \eta)^2} \sqrt{\frac{\tilde{A}(\eta) \tilde{B}(\eta)\tilde{D}(0)}{\tilde{A}(0) \tilde{D}^2(\eta) - \tilde{A}(\eta) \tilde{D}(\eta) \tilde{D}(0)}} \, .
\label{eq:hetaR}
\end{equation}
The new functions $\tilde{A}(\eta)$, $\tilde{B}(\eta)$, and $\tilde{D}(\eta)$ denote the metric functions with the substitution $r = R / (1 - \eta)$, with the fixed but unspecified chosen $R$.

This substitution provides two key advantages. First, it transforms the unspecified lower bound of the integral, $r=R$, to the fixed value $\eta = 0$. Second, the integration domain becomes finite: If the variable $r$ goes to $\infty$, as it does in our scenario if the observer position or the outer boundary of the disk is approached, $\eta$ goes to 1, thereby reducing computational difficulties and numerical errors.

Written in the variable $\eta$, Eqs.~\eqref{eq:firststep}, \eqref{eq:secondstepprimary}, \eqref{eq:secondstep} and  \eqref{eq:secondstepnoinversion} take the following form:
\begin{equation}
     \left(n'+\frac{1}{2} \right)\pi = 2 \int_0^1 h(\eta,R) \, d \eta \, ,
     \label{eq:firststep-eta}
\end{equation}
\begin{equation}
    \frac{\pi}{2} = \int_{\eta_{0 n'}}^1 h(\eta,R)\, d\eta \, ,
    \label{eq:secondstepprimary-eta}
\end{equation}
\begin{equation}
        \left( n+\frac{1}{2} \right)\pi = \int_0^{\eta_{n n'}} h(\eta,R) \, d \eta + \int_0^1 h(\eta,R) \, d \eta \, ,
    \label{eq:secondstep-eta}
\end{equation}
\begin{equation}
    \left( n+\frac{1}{2} \right) \pi = \int_{\eta_{n n'}}^1 h(\eta,R) \, d \eta \, .
     \label{eq:secondstepnoinversion-eta}
\end{equation}

\subsection{Merging with the shadow border --- limiting column}

An interesting special case arises when computing the inner radius $r_S^{\mathrm{in}}$ of a luminous disk for which the $n$-th image merges with all images of higher order. In particular, as discussed in Sec.~\ref{sec:matrix-merging}, since the values of the matrix of merging decrease in each row, this condition corresponds to the overlap between the image of order $n$ and the asymptotic image of order $n' \to \infty$, which approaches the boundary of the shadow. This situation occurs when the light ray defining the inner edge of the $n$-th image has an impact parameter equal to $b_{\mathrm{cr}}$, or equivalently, when the past-oriented light ray traced from $r_S$ spirals asymptotically toward the photon sphere. The corresponding limiting values of the radius of merging will be denoted as $r_{n\infty}$.

This limit can be computed by enforcing Eq.~\eqref{eq:secondstepnoinversion} with $R = r_{\mathrm{ph}}$:
\begin{equation} \label{eq:limitcondition-via-r}
\left( n+\frac{1}{2}\right)\pi = \int_{r_{n\infty}}^{+\infty} g(r,r_{\mathrm{ph}}) \, dr \ ,
\end{equation}
or
\begin{equation}
\left( n + \frac{1}{2} \right) \pi = \int_{\eta_{n\infty}}^{1} h(\eta, r_{\mathrm{ph}}) \, d \eta \ ,
\label{eq:limitcondition}
\end{equation}
where $\eta_{n\infty} = 1 - r_{\mathrm{ph}}/r_{n\infty}$.

The set of $r_{n\infty}$ values obtained in this way can be viewed as a column vector representing the limit of the matrix of merging:
\begin{equation}
r_{n\infty} = \lim_{n' \to \infty} r_{n n'} \, .
\end{equation}

\subsection{Step-by-step procedure for calculating the matrix of merging}
\label{subsec:step-by-step-procedure}

To summarize, we present here a step-by-step procedure for calculating the matrix of merging:
\begin{itemize}
\item Choose the specific spacetime metric by defining the metric coefficients in Eq.~\eqref{eq:metric}. The assumptions imposed on the metric are formulated in Subsec.~\ref{subsec:assumptions}.

\item Obtain the function $g(r, R)$ using Eq.~\eqref{eq:grR}.  

\item Compute the first row of the matrix, which contains the elements $r_{0 n'}$. For $n' = 1, 2, \ldots$, begin by solving Eq.~\eqref{eq:firststep} to determine the radius coordinate $R$ of the point of closest approach. This value is then to be substituted into Eq.~\eqref{eq:secondstepprimary} to calculate the corresponding element of the first row.  

\item Find the other entries $r_{n n'}$ with $n \neq 0$ by again solving Eq.~\eqref{eq:firststep} for $R$, and then applying Eq.~\eqref{eq:secondstep}. If this equation admits no solution --- indicating that the corresponding light ray does not reach a turning point --- Eq.~\eqref{eq:secondstepnoinversion} must be used instead. 

\item 
Find $r_{\mathrm{ph}}$ by solving Eq.~\eqref{eq:photon-sphere-equation} and selecting its outermost root. Next, for each $n=0,1,2,\ldots$ solve Eq.~\eqref{eq:limitcondition-via-r}; this yields the entries $r_{n\infty}$ in the limiting column of the matrix of merging.

\end{itemize}
If the variable $\eta$ is used, one must first obtain the functions $\tilde{A}(\eta)$, $\tilde{B}(\eta)$, $\tilde{D}(\eta)$, and write the function $h(\eta,R)$ according to \eqref{eq:hetaR}. Then, Eq.~\eqref{eq:firststep-eta} should be used instead of Eq.~\eqref{eq:firststep}; Eq.~\eqref{eq:secondstepprimary-eta} instead of Eq.~\eqref{eq:secondstepprimary};
Eq.~\eqref{eq:secondstep-eta} instead of Eq.~\eqref{eq:secondstep}; Eq.~\eqref{eq:secondstepnoinversion-eta} instead of Eq.~ \eqref{eq:secondstepnoinversion}; Eq.~\eqref{eq:limitcondition} instead of Eq.~\eqref{eq:limitcondition-via-r}.\\

\section{Example 1: Schwarzschild black hole}
\label{sec:example1-schw}

In the previous section, we explained in detail how to compute the matrix of merging numerically. In this and the next sections, we consider two examples: the Schwarzschild metric and the Reissner-Nordström metric.

The Schwarzschild metric coefficients can be written in the following form:
\begin{equation}
A(r) = B^{-1}(r) = 1 - \frac{2m}{r} \, , \qquad D(r)=r^2  \, ,
\end{equation}
where $m$ is the mass parameter with the dimension of a length.

In order to verify that our formalism is applicable to the Schwarzschild metric, we have to prove that in this spacetime for each order $n$ there is exactly one image. We will demonstrate in Sec.~\ref{sec:example2-RN} below that this is indeed true for the two-parameter family of Reissner-Nordstr{\"o}m black holes which includes the Schwarzschild black hole as a special case.

\begin{figure}
    \centering
    \includegraphics[width=0.49\textwidth]{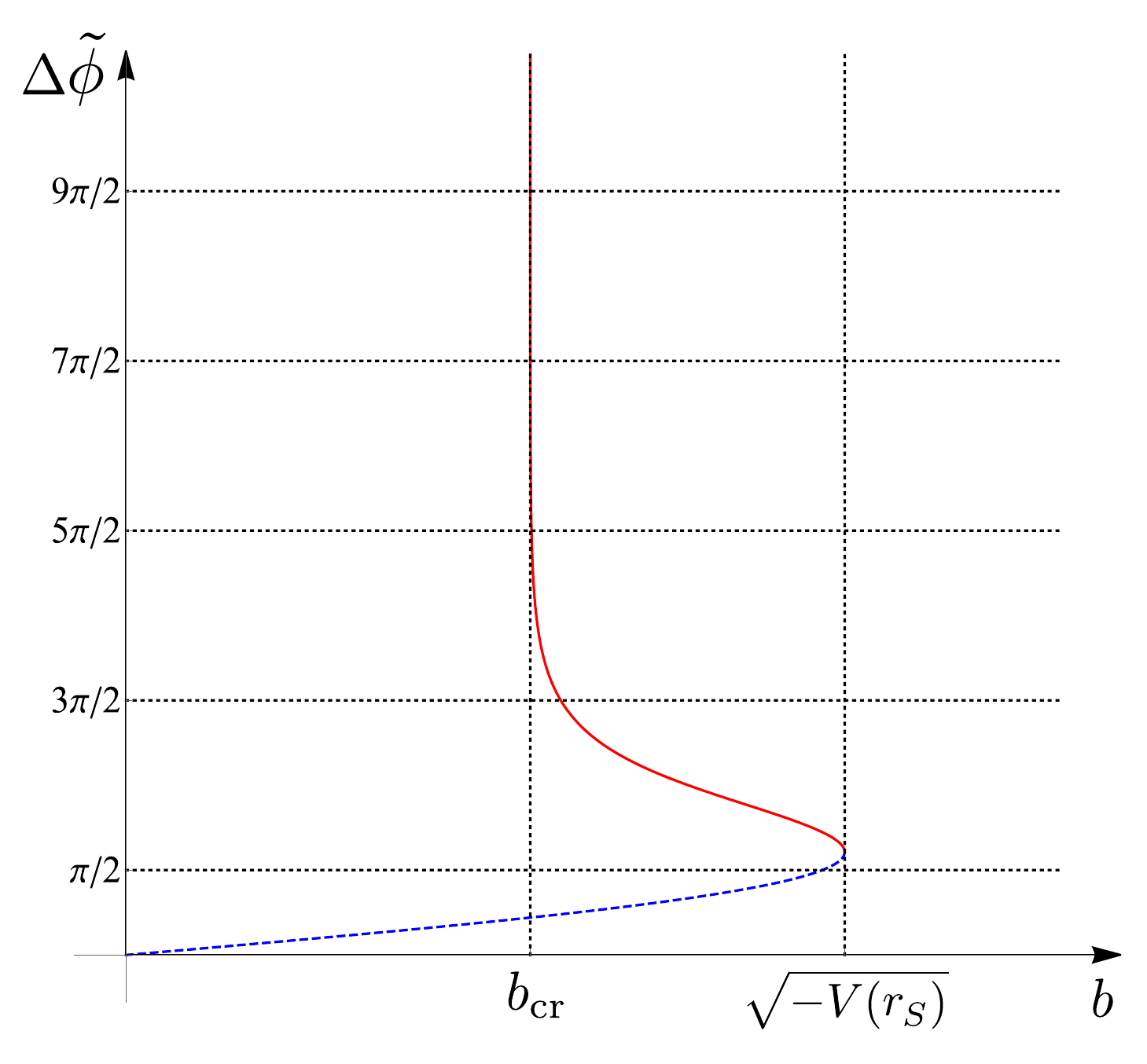}
    \caption{$\Delta \tilde{\phi}$ as a function of $b$ for the Schwarzschild spacetime with $r_S=8m$ and $r_O = \infty$. }
    \label{fig:Schwarzschild}
\end{figure}

Fig.~\ref{fig:Schwarzschild} illustrates that, indeed, in the Schwarzschild spacetime, there is exactly one image for each order $n$. The dashed (blue) part of the plot refers to light rays without turning points between the source and the observer. On this branch $b$ ranges from 0 (corresponding to a light ray that is radially outgoing) to $\sqrt{-V(r_S)}$ (corresponding to a light ray that leaves the light source at an angle of $90^\circ$ with respect to the radial direction). In the notation of Subsec.~\ref{subsec:calculation-light-deflection}, this branch corresponds to the case (ii), where (ii.a) is applicable when $b>b_{\mathrm{cr}}$ and (ii.b) is applicable when $b<b_{\mathrm{cr}}$. The solid (red) part of the plot refers to light rays that go through a turning point. On this branch $b$ ranges from $ b_{\mathrm{cr}}$ to $\sqrt{-V(r_S)}$. In terms of Subsec.~\ref{subsec:calculation-light-deflection}, this branch corresponds to the case (i). The fact that each horizontal line intersects the plot in exactly one point demonstrates that there is exactly one image for each order $n$. If the $\tilde{\phi}$ coordinates of the light source and the observer differ by $\pi/2$, as in the geometry we are considering, the intersection with the dotted line at $\Delta \tilde{\phi} = (n+1/2) \pi$ gives the impact parameter $b_n$ of the light ray that corresponds to the image of order $n$.

\subsection{Matrix of merging for Schwarzschild metric}

The explicit form of the functions introduced in the Eqs.~\eqref{eq:grR} and \eqref{eq:hetaR} can now be written:
\begin{equation}
    g(r,R)=\sqrt{\frac{R^3}{r^4(R-2m)-rR^3(r-2m)}} \, ,
\end{equation}
and
\begin{equation}
    h(\eta,R)=\sqrt{\frac{R}{\eta[-2m\eta^2+(6m-R)\eta+ 2(R-3m)]}}\, .
\end{equation}
Following the procedure described in Section \ref{sec:numerical}, we numerically calculate the values of the matrix elements $r_{n n'}$ and the column of the limiting values $r_{n\infty}$ through Eq.~\eqref{eq:limitcondition}:
\begin{widetext}
\begin{equation}
\begin{pmatrix}
& \quad  \; 5.24327 & 4.31621 & 4.28625 & 4.28498 & 4.28492 & \dots & 4.28492 \\
  &      & 3.27719 & 3.05332 & 3.04416 & 3.04377 & \dots & 3.04375\\
   &     &         & 3.22100 & 3.01092 & 3.00226 & \dots & 3.00187\\
    &    &         &         & 3.21862 & 3.00911 & \dots & 3.00008\\
     &   &         &         &         &  \vdots & \vdots & \vdots 
\end{pmatrix} \, .
\label{eq:Srmatrix}
\end{equation}
\end{widetext}
As we can notice, all $r_{n n'}$ values are below $6m$. This outcome supports the assertion made at the beginning of Sec.~\ref{sec:radius-merging} that when the inner boundary of the accretion disk lies at the innermost stable circular orbit (ISCO) radius, all images remain separated from each other.

\subsection{Analytical approximation for higher-order photon rings in the Schwarzschild metric}
\label{subsec:schw-SDL}

An interesting result, valid for images of order $n\geq 2$, can be obtained with the use of the so-called strong deflection limit \cite{darwin1959gravity, bozza2001g, bozza2002gravitational, bozza2007strong, Aldi-Bozza-2017}. As already mentioned in Sec.~\ref{sec:matrix-merging}, this technique provides an analytical approximation for the deflection angle near the divergence, which can be linked with the direction from which a photon arrives or, in other words, with the impact parameter \cite{bozza2002gravitational, Tsupko-2022-shape}.

In the Schwarzschild metric and for a polar observer the impact parameter for the image of order $n$ takes the following expression \cite{BK-Tsupko-2022, Tsupko-2022-shape}:
\begin{align}
    b_n = & \; 3\sqrt{3}m \biggl\{ 1 +
    \frac{6^5 \left(1 - 3m/r_S \right) }{ \left(3+\sqrt{3} \right)^2 } \nonumber \\ 
    &\times \left( 3+ \sqrt{3 + \frac{18m}{r_S}} \right)^{-2} 
    e^{- \left( n+\frac{1}{2} \right) \pi} \biggr\} \, ,
    \end{align}
or, equivalently \cite{Aratore2024},
\begin{align}
    b_n = & \; 3\sqrt{3}m \biggl\{ 1 +\frac{216}{2+\sqrt{3}}\left(1-\frac{3m}{r_S} \right) \nonumber \\ 
    &\times \left( 2 + \frac{3m}{r_S} + \sqrt{3+\frac{18m}{r_S}} \right)^{-1}
    e^{- \left( n+\frac{1}{2} \right)  \pi} \biggr\} \, ,
    \end{align}
Imposing the equality between $b_n$ for $r_S=r_{n n'}$ and $b_{n'}$ for $r_S \to +\infty$, we obtain an equation in which the only unknown is $r_{n n'}$ and the right hand side depends only on the difference between the orders $n$ and $n'$ of the considered images:
\begin{equation}
    \left(1 - \frac{3m}{r_{n n'}} \right)
    \left(2 + \frac{3m}{r_{n n'}} + \sqrt{3+\frac{18m}{r_{n n'}}} \right)^{-1} = k
\end{equation}
where we denote
\begin{equation}
    k = \frac{1}{2\sqrt{3}} e^{-(n'-n)\pi} \, .
\end{equation}
This equation can be solved analytically giving
\begin{equation}
    r_{n n'} = 3m \frac{(k+1)^2}{k^2-4k+1} \, .
    \label{eq:analyticrm}
\end{equation}\\
The resulting radius of merging $r_{nn'}$ depends only on the difference $(n' - n)$. This means that all diagonals starting from $n, n' \geq 2$ contain the same values --- of course, within the framework of this approximation only.

In particular, when $n'-n = 1$ we get $r_{n (n+1)} \approx 3.21852 \, m$ and when $n'-n=2$ we get $r_{n (n+2)} \approx 3.00902 \, m$. The agreement with the values calculated numerically in \eqref{eq:Srmatrix} is excellent: for example, $r_{2 3}$ is determined with a relative error of $8\times10^{-4}$ with respect to the numerically calculated value.\\

\section{Example 2: Reissner--Nordstr\"{o}m black hole}
\label{sec:example2-RN}

In the Reissner--Nordstr\"{o}m (RN) spacetime the metric is described by:
\begin{equation}
A(r) = B^{-1}(r) = 1 - \frac{2m}{r}+\frac{q^2}{r^2} \, ,
\quad D(r) = r^2  \, ,
\end{equation}
where $q$ is the electric charge in geometric units. Since for $q^2 > m^2$ the singularity becomes naked, we restrict ourselves to $q^2\leq m^2$. The spacetime is indeed asymptotically flat and the outermost photon sphere is at \cite{armenti1975existence, dadhich1977timelike, Howes-1981}
\begin{equation}
    r_{\mathrm{ph}} = \dfrac{3}{2} m + \sqrt{\dfrac{9m^2}{4} - 2 q^2} \, .
\label{eq:rphRN}
\end{equation}
We first have to prove that in the Reissner-Nordstr{\"o}m spacetime for each order $n$ there is exactly one image which is a necessary condition for our formalism to be applicable. We recall from  Subsec.~\ref{subsec:monotonicity} that this requires to prove that $\Delta \tilde{\phi}$ as given by (\ref{eq:Deltaphiofb}) is a monotonically decreasing function of $b$ on the interval $b_{\mathrm{cr}} < b < \sqrt{-V(r_S)}$. (On the other branch, where Eq.~\eqref{eq:Deltaphiofbnoinversion} is to be used, $\Delta \tilde{\phi}$ is always a monotonically increasing function of $b$, as we have proven in Subsec.~\ref{subsec:monotonicity}.)

To that end we write (\ref{eq:Deltaphiofb}) as
\begin{equation}
\Delta \tilde{\phi} = I_1-I_2-I_3 
\end{equation}
where
\begin{equation}
I_1 = \int _R ^{\infty}   
\sqrt{\frac{A(r)B(r)b^2}{D(r)^2-A(r)D(r)b^2}} \, dr \, ,
\end{equation}
\begin{equation}
I_2 = 
\int _{r_S} ^{\infty} 
\sqrt{\frac{A(r)B(r)b^2}{D(r)^2-A(r)D(r)b^2}} 
\, dr \, ,  
\end{equation}
\begin{equation}
I_3 = 
\int _{r_O} ^{\infty} 
\sqrt{\frac{A(r)B(r)b^2}{D(r)^2-A(r)D(r)b^2}}
\, dr \, .
\end{equation}
As $I_2$ and $I_3$ have the form of (\ref{eq:Deltaphiofbnoinversion}), we know that $dI_2/db > 0$ and $dI_3/db > 0$. What remains to be shown is that $dI_1 /db < 0$. To that end we rewrite $I_1$ as  
\begin{equation}
I_1 = \int _R ^{\infty}
\sqrt{\frac{A(r)B(r)D(R)}{D(r) \big(A(R)D(r)-A(r)D(R)\big)}}\, dr
\end{equation}
which for the Reissner-Nordstr{\"o}m metric takes the following form:
\begin{equation}
I_1 = \int _R ^{\infty}
\frac{R^2 \, dr }{\sqrt{r^4 ( R^2-2mR+q^2) - R^4 (r^2-2mr +q^2)}}\, .
\end{equation}
Substituting
\begin{equation}
z=\dfrac{R}{r} \, , \quad dz= - \dfrac{R}{r^2} dr
\end{equation}    
results in
\begin{equation}
I_1 = \int _0 ^1 \dfrac{R \, dz}{\sqrt{R^2(1 - z^2)-2mR (1-z^3)+q^2 (1-z^4)}}\, .
\end{equation}
With the substitutions
\begin{equation}
    z = \dfrac{R}{r} \, , \quad dz = - \dfrac{R}{r^2} dr \, ,
\end{equation}
this can be rewritten as
\begin{equation}
    I_1 =
    \int _0 ^1 \dfrac{R \, dz}{\sqrt{R^2(1-z^2) -2mR (1-z^3) +q^2 (1-z^4)}} \, ,
\end{equation}
hence
\begin{equation}
\dfrac{dI_1}{dR} 
= - \int _0 ^1 \dfrac{\Big(mR (1-z^3) -q^2 (1-z^4) \Big) \, dz}{\sqrt{R^2(1 - z^2)-2mR (1-z^3)+q^2 (1-z^4) \, }^3}\, .
\end{equation}
As $q^2 \le m^2 \le 2 m r_{\mathrm{ph}} /3< 2mR/3$, the integrand is manifestly positive on the entire interval $0 \le z < 1$ which proves that $dI_1/dR < 0$.
On the other hand, we find from (\ref{eq:VpR}) that 
\begin{equation}
    \dfrac{db}{dR} = \dfrac{R(R^2-3mR+2q^2)}{\sqrt{R^2-2mR+q^2 \,}^3} \, .
\end{equation}
With $R > r_{\mathrm{ph}}$ and (\ref{eq:rphRN}) this implies that
$db/dR > 0$, hence
\begin{equation}
 \dfrac{dI_1}{db} = \dfrac{dI_1}{dR} \dfrac{dR}{db} < 0
\end{equation}
which proves the desired monotonicity property.

\begin{widetext}

For the Reissner-Nordstr{\"o}m metric, the functions defined in Eqs. \eqref{eq:grR} and \eqref{eq:hetaR} become:
\begin{equation}
    g(r,R)=\sqrt{\frac{R^3}{r^4(q^2-2mR+R^2)-R^4(q^2-2mr+r^2)}}\, ,
\end{equation}
\begin{equation}
    h(\eta,R)=\sqrt{\frac{R^2}{z\left[(q^2z^3-2(2q^2-mR)z^2+(6q^2-6mR+   R^2)z-4q^2+6mR-2R^2)\right]}}\, .
\end{equation}

If $q=0.5 \, m$, the matrix of merging reads:
\begin{equation}
\begin{pmatrix}
& \quad \; 5.07116 & 4.11768 & 4.08410 & 4.08252 & 4.08245 & \dots & 4.08244\\
       & & 3.10920 & 2.88039 & 2.87011 & 2.86962 & \dots & 2.86960\\
       & &         & 3.04866 & 2.83524 & 2.82557 & \dots & 2.82508\\
       & &         &         & 3.04583 & 2.83310 & \dots & 2.82298\\
       & &         &         &         &  \vdots & \vdots & \vdots
\end{pmatrix} \, ,
\label{eq:RN05rmatrix}
\end{equation}
while for $q=m$, it is:
\begin{equation}
\begin{pmatrix}
& \quad \; 4.45930 & 3.30547 & 3.22830 & 3.22040 & 3.21955 & \dots & 3.21945\\
   &     & 2.41409 & 2.11812 & 2.09061 & 2.08768 & \dots & 2.08732\\
    &    &         & 2.30297 & 2.03733 & 2.01214 & \dots & 2.00912\\
     &   &         &         & 2.29179 & 2.02896 & \dots & 2.00098\\
      &  &         &         &         &  \vdots & \vdots & \vdots
\end{pmatrix} \, .
\label{eq:RN1rmatrix}
\end{equation}
\end{widetext}

As the most interesting result, we observe a major difference between a weakly charged and a strongly charged Reissner-Nordstr{\"o}m black hole if we assume that the radius coordinate of the inner boundary of the disk coincides with the ISCO radius. The latter is the unique real solution of the cubic equation 
\begin{equation}
    r^3-6mr^2+9q^2r-4q^4/m = 0 \, ,
\end{equation}
see e.g. \cite{armenti1975existence, dadhich1977timelike, Howes-1981}.
As for $q = 0.5 \, m$ the ISCO radius is approximately at $5.6 \, m$, we read from the matrix in (\ref{eq:RN05rmatrix}) that the images are separated. For $q = m$, however, the ISCO radius is at $4 \, m$ which is smaller than the value of $r_{0 1}$ according to the matrix \eqref{eq:RN1rmatrix}. This means that, considering an accretion disk that extends down to the ISCO, there must be a critical value $q_m$ of the charge parameter, somewhere in the interval $0.5 \, m < q_m < m$, such that a Reissner-Nordstr{\"o}m black hole with charge $q_m < q \le m$ will have the primary image ($n=0$) already overlapped with the secondary one ($n=1$). Numerically we have found that $q_m \approx 0.853 \, m$.

\section{Constraining the spacetime metric or the accretion model through the overlapping pattern of photon rings}
\label{sec:constraining}

In this Section, we discuss how observations of photon
ring overlaps can be used to probe either the spacetime metric or the accretion model. In our simplified framework, the accretion model is fully specified by the inner radius of the luminous disk, $r_S^{\mathrm{in}}$. If this inner radius is known --- for instance, if it coincides with the innermost stable circular orbit (ISCO) --- then the observed pattern of overlapping photon rings can serve as a probe of the underlying spacetime metric. Conversely, if the metric is known, the observed overlap pattern allows us to estimate the inner radius of the luminous disk, thereby placing constraints on the accretion model (compare, e.g., with Ref.~\cite{Kocherlakota-Rezzolla-2022}).

In Sec.~\ref{sec:numerical}, we presented the numerical procedure used to compute the values of $r_S^{\mathrm{in}}$ at which two photon rings merge. Numerical examples for specific metrics were provided in Sections \ref{sec:example1-schw} and \ref{sec:example2-RN}. In the present section, we fix different values of the inner edge of the accretion disk, $r_S^{\mathrm{in}}$, and for each choice we investigate the resulting overlapping pattern of photon rings --- that is, which images appear merged to a distant observer.

The figures in this section show the locations of the first three photon rings, computed numerically for a chosen value of $r_S^{\mathrm{in}}$. Notably, the question of which photon rings overlap and which remain separated can, in fact, be addressed without performing such numerical calculations. It is sufficient to specify the inner radius and compare it with the corresponding values in the matrix of merging. This matrix alone encodes the complete qualitative picture of the overlap pattern, making it a useful diagnostic tool for interpreting observational patterns.

\subsection{Numerical calculations of images}

In this subsection, we describe the numerical procedure for calculating the images. Many related aspects have already been discussed in Sec.~\ref{sec:numerical}, where the calculation of radii of merging was considered. Here, however, we also address source positions $r_S$ that produce parts of images inside the shadow and, correspondingly, light rays with $b < b_\mathrm{cr}$. Therefore, additional details must be provided for clarity.

To characterize the angular positions of the images, we compute the associated impact parameters, which are related to the observed angles via Eq.~\eqref{eq:theta-b-rO}. In our model, the disk extends from $r_S^{\mathrm{in}}$ to infinity. As a result, the impact parameter corresponding to the primary image is unbounded above, whereas those for higher-order images are finite.

Consider first the primary image ($n=0$). In this case, photons do not go through a radial inversion point, and Eq.~\eqref{eq:Deltaphiofetanoinversion} must be used:
\begin{equation} \label{eq:prim-image-calc-01}
\frac{\pi}{2} = \int_{\eta_S}^{1} h(\eta, R) \, d \eta \, .
\end{equation}
This equation is solved numerically to determine $R$, from which the impact parameter can be obtained via Eq.~\eqref{eq:impactpar}. This yields the angular location of the inner edge of the primary image.

Complications arise when $r_S < r_{0 \infty}$, where $r_{0\infty}$ is the limiting radius for the primary image computed from Eq.~\eqref{eq:limitcondition}. In this regime, the backward-traced light ray approaches the horizon and Eq.~\eqref{eq:prim-image-calc-01} returns a complex value for $R$. In such cases, Eq.~\eqref{eq:Deltaphiofbnoinversion} must be used instead:

\begin{equation}
\frac{\pi}{2} = \int_{r_S^{\mathrm{in}}}^{+\infty} f(r, b) \, dr \, .
\end{equation}
Solving this equation directly yields the impact parameter $b$. In this situation, one finds $b < b_{\mathrm{cr}}$, implying that the inner edge of the image lies inside the black hole shadow.

For images of higher orders $n=1,2,...$, the same logic applies, but with the corresponding deflection:
\begin{equation}
\left( n + \frac{1}{2} \right) \pi = \int_{0}^{\eta_S} h(\eta, R) \, d\eta + \int_{0}^{1} h(\eta, R) \, d \eta \, .
\label{eq:determingRgivenrS}
\end{equation}
This equation corresponds to case (i) from Subsec.~\ref{subsec:calculation-light-deflection}.
For the outer boundaries of images, the light ray always passes through a turning point, so Eq.~\eqref{eq:determingRgivenrS} always applies. Setting $r_S \to \infty$ (i.e., $\eta_S \to 1$) gives the corresponding impact parameter $b$.

For calculation of the inner boundaries of images, one substitutes $r_S^{\mathrm{in}}$ into Eq.~\eqref{eq:determingRgivenrS}.
If this equation admits no solution for $R$, then we must switch to case (ii) in Subsec.~\ref{subsec:calculation-light-deflection} and compute the deflection using Eq.~\eqref{eq:Deltaphiofbnoinversion}. Note that in this calculation it is not necessary to distinguish between cases (ii.a) and (ii.b): since our goal is to determine $b$, the same formula can be applied to both. Accordingly, we write
\begin{equation}
\left( n + \frac{1}{2} \right) \pi = \int_{r_S^{\mathrm{in}}}^{+\infty} f(r, b) \, dr \, ,
\end{equation}
and solve for $b$.

In this way, we obtain the impact parameter for the inner border of the primary image, as well as the list of impact parameters corresponding to the inner and outer borders of each image as seen by the distant observer. Additionally, in all plots that follow, the shadow radius is indicated by a black dot.

\subsection{Testing the spacetime metric through photon ring overlaps}

\begin{figure}[ht]
    \centering
    \includegraphics[width=0.48\textwidth]{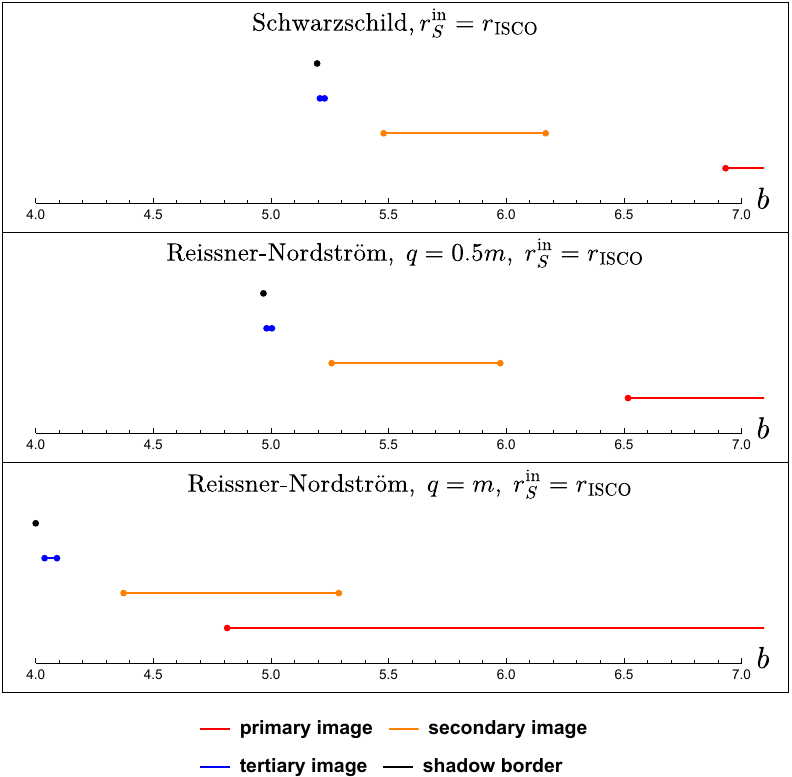}
    \caption{Photon rings produced by a luminous accretion disk around different types of black holes, with the inner boundary of the disk placed at the ISCO radius for the corresponding spacetime metric. Each panel shows the first three photon rings in specific metric: the $n=0$ ring (primary image), $n=1$ ring (secondary image), and $n=2$ ring (tertiary image). The black dot indicates the edge of the black hole shadow. In the upper and middle panels, corresponding to the Schwarzschild and Reissner--Nordström metric with intermediate charge, all presented rings are clearly separated. In contrast, the lower panel, corresponding to the extreme RN black hole, shows an overlap between the primary and secondary images. This qualitative difference implies that, under the assumption that the accretion disk has its inner edge at the ISCO, the extreme RN black hole is observationally distinguishable from both the Schwarzschild and the moderately charged RN cases.}
    \label{fig:LineplotISCO}
\end{figure}

As outlined above, a prescribed accretion model allows one to probe the underlying spacetime metric through the observed overlaps of photon rings. In the simple model used in this paper, the accretion disk is completely determined by one parameter, namely the radius coordinate of the inner boundary, $r_S^{\mathrm{in}}$. In this section we consider a scenario in which the inner boundary of the accretion disk coincides with the ISCO radius, see Fig.~\ref{fig:LineplotISCO}.

In the Schwarzschild metric, the ISCO radius is $r_\mathrm{ISCO}=6m$  \cite{Kaplan-1949}. This value exceeds the radius of merging $r_{01}$ obtained in Eq.~\eqref{eq:Srmatrix} (and, consequently, all other elements of the matrix). Therefore, all photon rings remain separated from one another. A similar qualitative behavior occurs in the RN metric with charge $0 \le q \lesssim 0.853 \, m$, recall Subsec.~\ref{sec:example2-RN}. This, however, is not the case for $0.853 \, m \lesssim q \le m$. In particular, in the extreme RN metric, $q=m$, the ISCO radius is reduced to $r_\mathrm{ISCO}=4m$, which is smaller than $r_{01}$ in \eqref{eq:RN1rmatrix}. As a result, in this case the overlap between primary and secondary images is found (Fig.~\ref{fig:LineplotISCO}).

Thus, under the assumption that the accretion disk extends to the ISCO, this observable difference provides a way to distinguish between different black hole metrics. In particular, an overlap between the primary and secondary image would rule out Schwarzschild and moderately charged  RN black holes, favoring the strongly charged RN case.

\begin{figure}[ht]
    \centering
    \includegraphics[width=0.48\textwidth]{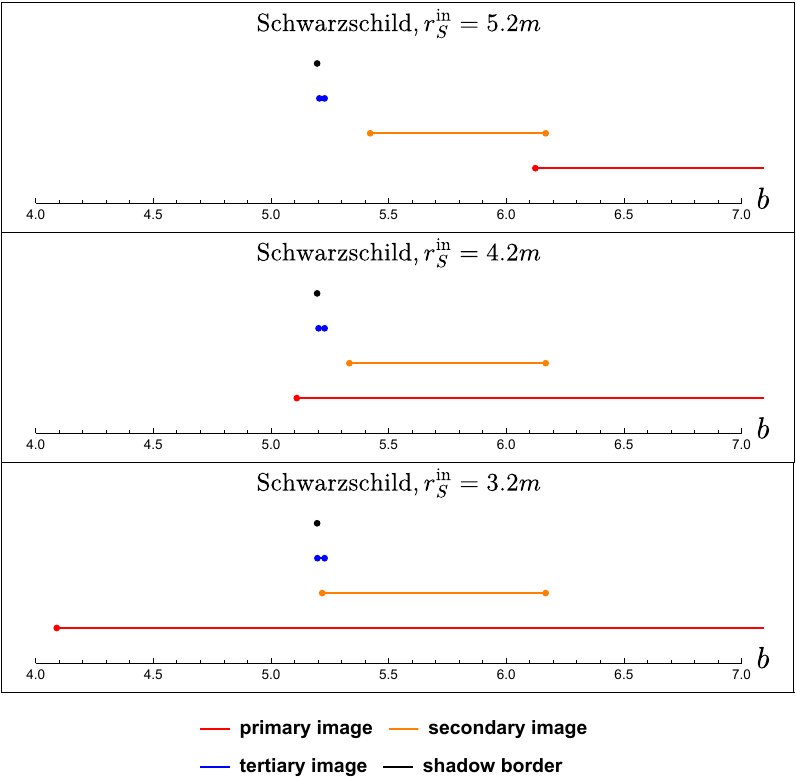}    \caption{Photon rings around a Schwarzschild black hole for different values of the disk inner radius: $5.2m$, $4.2m$, and $3.2m$. For the case with $r_S^{\mathrm{in}} = 6m$, where all images are clearly separated, see the upper panel of Fig.~\ref{fig:LineplotISCO}. As the disk inner radius decreases, the primary image first overlaps with the secondary image (upper panel), then with the tertiary image (middle panel). Only after further decrease, the secondary and tertiary images begin to overlap (lower panel). This demonstrates that there cannot exist an overlapping pattern in which the secondary and tertiary images overlap without both of them overlapping with the primary image. As shown in Sec.~\ref{sec:matrix-merging}, this property is universal for all spherically symmetric metrics.}
    \label{fig:LineplotShw}
\end{figure}

\begin{figure}[ht]
    \centering
    \includegraphics[width=0.48\textwidth]{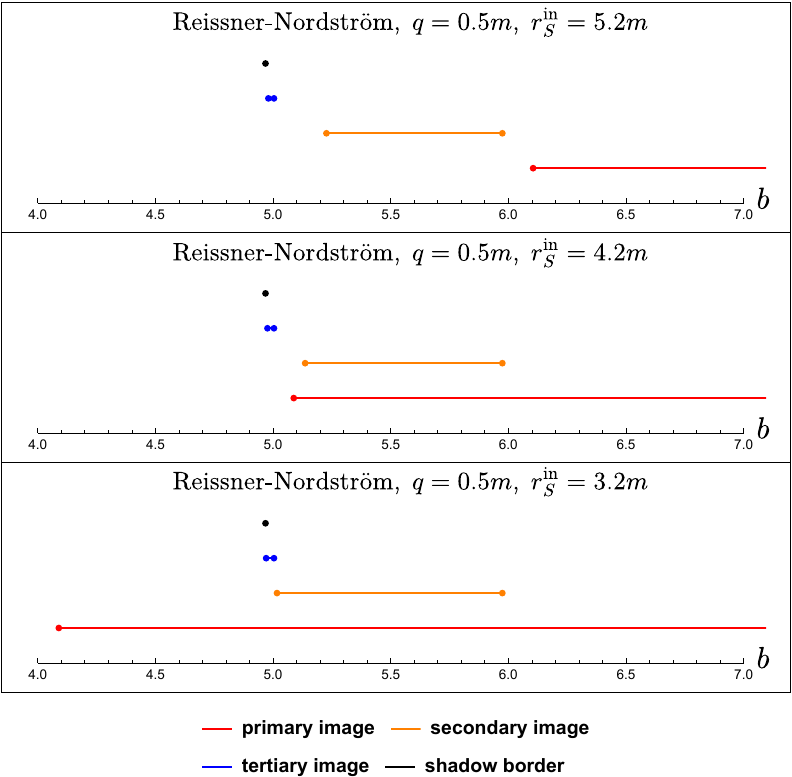}
    \caption{Photon rings in the Reissner--Nordstr\"om metric with $q=0.5m$ for different values of the disk inner radius. The values of the inner radius are the same as in Fig.~\ref{fig:LineplotShw}.}
    \label{fig:LineplotRN05}
\end{figure}

\begin{figure}[ht]
    \centering
    \includegraphics[width=0.48\textwidth]{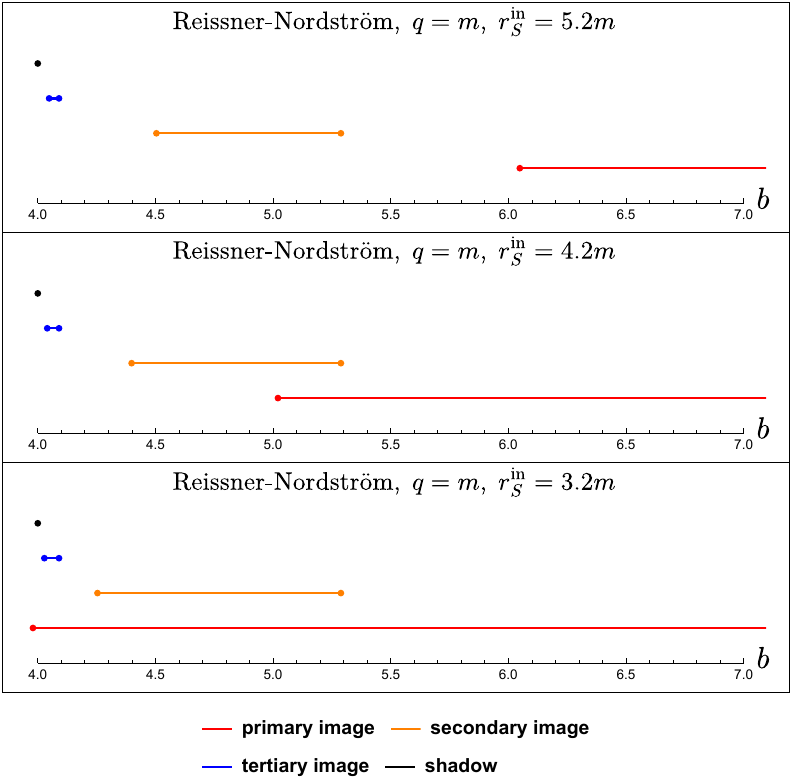}
    \caption{Photon rings in the Reissner--Nordstr\"om metric with $q=m$ for different values of the disk inner radius. The values of the inner radius are the same as in Fig.~\ref{fig:LineplotShw} and Fig.~\ref{fig:LineplotRN05}.}
    \label{fig:LineplotRN1}
\end{figure}

\subsection{Constraining the accretion model through photon ring overlaps}

In this subsection, we reverse the logic compared to the previous one: we assume that the spacetime metric is known and vary the inner boundary of the accretion disk, which in our simplified model is the only distinguishing feature of the accretion flow. Our aim is to demonstrate that the observed overlapping pattern of photon rings can place meaningful constraints on the accretion model.

In each of the Figs.~\ref{fig:LineplotShw}, \ref{fig:LineplotRN05}, and \ref{fig:LineplotRN1}, the black hole metric is fixed, while the photon ring structure is shown for decreasing values of $r_S^\mathrm{in}$. The specific values of the inner radius are chosen to better visualize the qualitative differences between the cases.

In Fig.~\ref{fig:LineplotShw}, we consider the Schwarzschild metric. The upper panel shows $r_S^{\mathrm{in}} = 5.2 \, m < r_{01}$, resulting in an overlap between the primary and secondary images, while higher orders remain separated. In the central panel, $r_S^{\mathrm{in}} = 4.2 \, m$ is smaller than the limiting radius $r_{0 \infty}$ in the matrix \eqref{eq:Srmatrix}. The primary image now extends inside the shadow and overlaps all images of higher order.

It is important to recall that light rays approaching the compact object (i.e., moving inward) with the critical impact parameter $b_{\mathrm{cr}}$ are injected in an unstable circular orbit, and only photons with $b > b_{\mathrm{cr}}$, after passing through a turning point, can reach the observer. If the impact parameter is smaller than the critical value, the photon inevitably plunges into the singularity. As a consequence of this behavior, an observer in the asymptotically flat region perceives the black hole shadow with angular radius $\theta_{\mathrm{cr}} = b_{\mathrm{cr}}/r_O$. However, if a light source is located near the black hole, some photons can still reach the observer even with $b < b_{\mathrm{cr}}$, provided they move outward from the source without inverting their radial motion [case (ii.b) of Subsec.~\ref{subsec:calculation-light-deflection}]. Such photons make parts of the shadow appear bright --- see the more detailed discussion in Subsec.~\ref{subsec:images-inside-shadow}.

In the bottom panel of Fig.~\ref{fig:LineplotShw}, $r_S^{\mathrm{in}} = 3.2 \, m < r_{12}$, leading to overlap between the secondary and tertiary images. Actually, from the analytic approximation obtained via the strong deflection limit Eq.~\eqref{eq:analyticrm}, this value of $r_S^{\mathrm{in}}$ is also smaller than all radii of merging $r_{n(n+1)}$, implying that all higher-order images (if drawn) would overlap with their immediate neighbors.

Fig.~\ref{fig:LineplotRN05} (RN with $q=0.5m$) shows similar qualitative trends: distinct images in the top panel; full overlap of primary and secondary in the middle; and a primary image that dips below the shadow boundary in the bottom, without further overlaps.

Finally, Fig.~\ref{fig:LineplotRN1} (extreme RN) displays partial overlap between the primary and secondary images in the middle panel, and a primary image just entering the shadow in the lower panel.

An illustrative example of how different accretion models affect the appearance of higher-order images for the same spacetime metric is presented in Figure 5 of Kocherlakota \textit{et al.} \cite{Kocherlakota-2024b}. Using an adaptive scale, the authors clearly display images up to order $n=4$. They present two different accretion models in the Schwarzschild black hole spacetime. The first model features an equatorial thin disk with an inner boundary at the ISCO radius ($6m$) and a very large outer boundary. In this case, all the images shown are clearly separated. In contrast, the second model assumes a spherical emission region extending all the way to the horizon $(2m)$, resulting in all the shown images being overlapped.\\

\section{Summary and conclusions}
\label{sec:conclusions}

(i) In this paper, we investigate the overlapping of photon rings --- higher-order lensed images of a black hole’s luminous environment, concentrated near the shadow boundary and expected to be detected in future observations. We work within a broad class of static, spherically symmetric spacetimes. The specific conditions (reasonably general) imposed on the metric are detailed in Subsec.~\ref{subsec:assumptions}. We consider an idealized model of accretion, represented by a geometrically thin equatorial disk with specified inner and outer radii, and an observer located on the symmetry axis (see Subsec.~\ref{subsec:photon-rings} and Fig.~\ref{fig:merging}). To further simplify the analysis, we consider the limit that the outer radius of the disk is at infinity. Under this assumption, for any fixed metric the question of overlapping is entirely determined by the inner radius of the accretion disk (Subsec.~\ref{subsec:overlap}).

(ii) Depending on the value of the inner radius of the accretion disk, the thickness of each photon ring varies, and therefore they may or may not overlap. (By overlapping, we mean that portions of images appear at the same angular position on the observer’s sky.) When the inner radius is sufficiently large, the photon rings are thin and separated from one another. If it is smaller, the rings become thicker, and the separations between them disappear. For even smaller values, more and more rings overlap.

To characterize this behavior, we introduce the radius of merging --- defined as the value of the inner radius of the accretion disk at which the photon rings of two given orders begin to overlap (Subsec.~\ref{subsec:radius-merging-def}). An illustrative example involving the primary and secondary images is shown in Fig.~\ref{fig:merging}.

(iii) Since each radius of merging is labeled by two indices corresponding to the image orders, we find it convenient to arrange the radii of merging in the scheme of a matrix and to analyze the properties of overlapping within this matrix-based framework. We refer to it as the matrix of merging (Sec.~\ref{sec:matrix-merging}). The matrix of merging is assumed to be calculated numerically for each spacetime metric under consideration. Once the elements of the matrix are known, they allow one to fully describe the overlapping pattern qualitatively for any chosen value of the accretion disk’s inner boundary.

(iv) A remarkable feature of the matrix of merging is that it has several universal properties that are identical across all metrics of the considered class. Moreover, these properties can be identified without explicitly calculating the light ray trajectories. These properties are collected in Sec.~\ref{sec:matrix-merging}. In particular, we show that, with decreasing inner radius of the luminous disk, first the primary ($n=0$) image will intersect with the secondary ($n=1$) and tertiary ($n=2$) image, and only then the secondary ($n=1$) image will intersect with the tertiary ($n=2$) image; see, e.g., Fig.~\ref{fig:LineplotShw}. In other words, for any spherically symmetric metric and any choice of the disk's inner radius, it is not possible to observe an overlapping pattern in which the secondary and tertiary images overlap without the primary image overlapping with both of them. Realizable and forbidden overlapping patterns of the first three images are shown in Fig.~\ref{fig:forbidden}.

(v) We present a detailed discussion of the calculation of light deflection and radii of merging, see Sections \ref{sec:light-deflection} and \ref{sec:numerical}, respectively. In Sections \ref{sec:example1-schw} and \ref{sec:example2-RN}, we present the merging matrices for two cases found numerically: the Schwarzschild black hole and the Reissner–Nordström black hole.

(vi) The main application of our study of overlapping is the analysis of the spacetime metric or accretion model, based on the observed merging or separations of photon rings of specific orders (Sec.~\ref{sec:constraining}). The matrix-based framework allows this analysis to be carried out in a more systematic and structured way.

An interesting finding we would like to highlight is that, when the accretion disk has its inner boundary at the ISCO radius, the weakly charged and the strongly charged Reissner--Nordstr{\"o}m black holes exhibit qualitatively different overlapping patterns. In particular, these patterns are different for the Schwarzschild metric and the extreme Reissner--Nordstr{\"o}m metric, see Fig.~\ref{fig:LineplotISCO}. In the Schwarzschild case, the primary and secondary images remain separated, whereas in the extreme Reissner--Nordström case, they overlap.

(vii) The overlapping of the $n=0$ and $n=1$ photon rings can already be studied with observations from forthcoming projects such as the next generation Event Horizon Telescope \cite{Johnson-2023-Galaxies, Ayzenberg-2025-review} and the Black Hole Explorer \cite{Johnson-2024-BHEX}. These projects will probe secondary images around M87* and Sgr A*. A further increase in observational resolution is expected to enable resolving the $n=1$ photon rings of other supermassive black holes and, with even higher resolution, the $n=2$ photon rings \cite{pesce2021toward}. Accordingly, our results regarding the overlapping of $n=0$, $n=1$, and $n=2$ photon rings would become applicable.

We emphasize that the definition of overlapping and the derived properties and criteria are purely geometric, as they rely on the coincidence of the ring boundaries. They are independent of emissivity or radiative transfer, which may affect observational appearance. In particular, we don't calculate intensities of the images. Additionally, in our model, we assume the luminous disk to have an infinite outer extent. In practice, the emission region is finite, and the presence or absence of overlapping also depends on the position of the outer boundary. Nevertheless, since our analysis covers all possible emission radii in the disk's outer regions, many of our results can be applied to disks of finite outer size as well. For example, if our model with an infinitely extended disk predicts no overlapping, the same conclusion holds for any disk with a finite outer boundary.

(viii) There are a number of directions in which our study can be extended. First, since our method is formulated for a large class of spherically symmetric and static spacetimes, it is straight-forward to calculate the matrix of merging for metrics other than the Schwarzschild or Reissner--Nordström cases considered above. A step-by-step procedure for calculatiing this matrix is outlined in Subsec.~\ref{subsec:step-by-step-procedure}. Second, the analysis can be generalized to axisymmetric spacetimes, such as the Kerr black hole, which is also straight-forward as long as we consider an accretion disk in the equatorial plane and a polar observer. Third, similar consideration can be performed in the presence of a surrounding medium, for example a cold plasma, which is a dispersive medium. Then the images of each order would depend on the frequency.\\

\section*{Acknowledgments}

F.A. thanks Eva Hackmann for the kind hospitality at ZARM, University of Bremen. O.Y.T. thanks the German Research Foundation (Deutsche Forschungsgemeinschaft, DFG) under Germany's Excellence Strategy --- EXC-2123 QuantumFrontiers --- 390837967 for supporting his visit to ZARM, University of Bremen, and Eva Hackmann and her group for their kind hospitality and useful discussions.


%

\end{document}